# Structural Identifiability of Cyclic Graphical Models of Biological Networks with Latent Variables

Yulin Wang[1], Na Lu[2], Hongyu Miao[3,*]

[1] *School of Computer Science and Engineering, University of Electronic Science and Technology of China, Chengdu, Sichuan, China*

[2] *State Key Laboratory for Manufacturing Systems Engineering, Systems Engineering Institute, Xi'an Jiaotong University, Xi'an, Shaanxi, China*

[3] *Department of Biostatistics, School of Public Health, University of Texas Health Science Center at Houston, Houston, TX, 77030, USA*

## Abstract

**Background:** Graphical models have long been used to describe biological networks for a variety of important tasks such as the determination of key biological parameters, and the structure of graphical model ultimately determines whether such unknown parameters can be unambiguously obtained from experimental observations (i.e., the identifiability problem). Limited by resources or technical capacities, complex biological networks are usually partially observed in experiment, which thus introduces latent variables into the corresponding graphical models. A number of previous studies have tackled the parameter identifiability problem for graphical models such as linear structural equation models (SEMs) with or without latent variables. However, the limited resolution and efficiency of existing approaches necessarily calls for further development of novel structural identifiability analysis algorithms.

**Results:** An efficient structural identifiability analysis algorithm is developed in this study for a broad range of network structures. The proposed method adopts the Wright's path coefficient method to generate identifiability equations in forms of symbolic polynomials, and then converts these symbolic equations to binary matrices (called identifiability matrix). Several matrix operations are introduced for identifiability matrix reduction with system equivalency maintained. Based on the reduced identifiability matrices, the structural identifiability of each parameter is determined. A number of benchmark models are used to verify the validity of the proposed approach. Finally, the network module for influenza A virus replication is employed as a real example to illustrate the application of the proposed approach in practice.

**Conclusions:** The proposed approach can deal with cyclic networks with latent variables. The key advantage is that it intentionally avoids symbolic computation and is thus highly efficient. Also, this method is capable of determining the identifiability of each single parameter and is thus of higher resolution in comparison with many existing approaches. Overall, this study provides a basis for systematic examination and refinement of graphical models of biological networks from the identifiability point of view, and it has a significant potential to be extended to more complex network structures or high-dimensional systems.

## Background

Although the reductionism approaches have led to tremendous success in advancing our knowledge and understanding of individual biological components and their functions, it has been broadly recognized that many organic/cellular functions or disorders cannot be attributed to an individual molecule [1]. Instead, numerous biological components interact with each other and orchestrate various dynamic events that are critical to the beginning and extension of life [2]. To systematically investigate and understand such complex interactions, a variety of biological networks (e.g., transcriptional and post-transcriptional regulatory networks [3-6], functional RNA networks [7-9], protein-protein interaction networks [10, 11], and metabolic networks [12, 13]) have necessarily been constructed based on experimental observations or predictions. Nowadays, biological networks are playing critical roles in biomedical research and practice at multiple levels or scales (e.g., genetics [14], immunology [15], cancer [16], drug discovery [17, 18]), and the associated modeling and computation techniques and tools are under active development for network property investigation, network structure identification, experimental data analysis and interpretation, and so on [1, 15-19].

Graphical models are one of the most powerful mathematical languages for biological network representation, and have long been used for various quantitative analysis tasks [19-21]. In particular, the determination of unknown model parameter values from experimental data is of fundamental importance to many other critical tasks (e.g., computer simulation or prediction, network structure refinement), and it should be stressed that parameter identifiability is one of the first questions that needs to be answered before any statistical method can be applied to obtain accurate and reliable estimates of unknown parameters [20]. More specifically, limited by resources or technical capabilities, it is not uncommon that only part of the nodes or interactions (i.e., edges) in a biological network can be experimentally observed such that the values of certain unknown parameters associated with those unobserved nodes or edges cannot be uniquely determined from experimental data due to the lack of information. However, even if all the nodes and edges are observed, identifiability issues may still occur due to, e.g., model misspecification. It is thus necessary to develop identifiability analysis techniques for graphical models with or without latent variables.

Since graphical models refer to a broad range of mathematical formulations [19-22], it is impossible to explore the identifiability analysis techniques for all different types of graphical models in one study. Here we focus on the structural identifiability analysis problem of static linear structural equation model (SEM), which is a representative and generic graphical model type that has been widely used in many different research areas such as clinical psychology, education, cognitive science, behavioral medicine, developmental psychology, casual inference [23, 24], and systems biology [25-27]. A number of previous studies have proposed identifiability analysis techniques for linear SEMs with or without latent variables [23, 24, 28-43]. More specifically, the traditional method described in [23] constructs a so-called system matrix from a given model structure and derives the rank and order conditions based on this matrix for identifiability analysis. However, this approach can only handle comparatively simple network structures (e.g., block recursive models [23]) without latent variables, and cannot deal with the disturbance correlation between variables (i.e., nodes). To deal with a broader range of model structures, investigators from different disciplines have made further attempts by considering the topological or other features of certain networks. For instance, several previous studies have derived the sufficient criteria for parameter identifiability based on local characteristics of subnetworks, including Pearl's back door and front door criteria [24], Brito and Pearl's generalized instrumental variable criterion [30], and Tian's accessory set approach [41]. For certain network structures, sufficient conditions for parameter identifiability have also been established for the entire network instead of subnetworks; e.g., Brito and Pearl's conditions for bow-free models [28], Brito and Pearl's auxiliary sets condition for directed acyclic graph (DAG) models [36], Drton's condition based on injective parametrization of mixed graphs [35], and Foygel's half-trek criterion for mixed graphs [37]. While the criteria and conditions mentioned above are important progresses made in the field, they only provide a partial or overall assessment of parameter identifiability.

To determine the identifiability of every single parameter in the model, Tian [32] adopted the partial regression analysis technique, but this approach can only handle a special class of P-structure-free SEMs. Also, Sullivant et al. [34] tackled this problem using a computer algebra method, which turns out to be applicable only to SEMs with a small number of variables due to the prohibitive computation costs associated with Gröbner basis reduction. Therefore, it is still necessary to develop more efficient single-parameter-level approaches for structural identifiability analysis of whole networks.

In this study, we developed a novel and efficient approach for structural identifiability analysis of cyclic linear SEMs with latent variables. The proposed method is applicable to both directed cyclic and acyclic graphs with or without latent variables, and thus presents an extension of existing algorithms in terms of generality. Different from other existing algebraic approaches, although our method uses the Wright's path coefficient method to generate identifiability equations in forms of nonlinear symbolic polynomials, it avoids the expensive symbolic computations (e.g., Gröbner basis reduction) by converting identifiability equations to binary matrices, and is thus highly efficient. Moreover, in contrast to other methods that can only draw conclusions on the overall identifiability of a model, the proposed method can determine the identifiability of each single unknown parameter, and is thus of higher resolution and enables researchers to locate the problematic subnetwork structures to refine model structures or improve experimental design. We collected a number of benchmark models from literature and verified the validity of our method using those models. Finally, we applied our method to the network module for influenza A virus (IAV) within-host replication to gain insights into parameter identifiability and experimental design.

# Methods

The key definitions and steps involved in the proposed algorithm are described in this section, including the definition of structural identifiability analysis for cyclic SEMs, the generation of identifiability equations, the conversion to identifiability matrices, and the symbolic-free identifiability determination based on the reduced identifiability matrices. The necessary theoretical justification is also given.

### SEM and structural identifiability

The structural equation models considered in this study correspond to a mixed cyclic graph $G = (\mathbf{V}, \mathbf{D}, \mathbf{U})$, where $\mathbf{V}$ is a set of vertices, $\mathbf{D}$ a set of directed edges, and $\mathbf{U}$ a set of undirected edges. That is, in the SEM, each model variable $Y_i$ corresponds to a vertex $V_i$ ($i = 1, 2, \ldots, n$), the structure of the coefficient matrix $\mathbf{C} = [c_{ij}]$ is specified by $\mathbf{D}$ (i.e., $c_{ij}$ exists if a directed edge from $V_j$ to $V_i$ is in $\mathbf{D}$; otherwise, $c_{ij} = 0$ if no edge exists in $\mathbf{D}$ from $V_j$ to $V_i$, $i \neq j$), and the existence of disturbance correlation between two variables is given by $\mathbf{U}$. Here disturbance refers to all the omitted causes of a variable, and disturbance correlation is the correlation between two variables due to the existence of common omitted cause(s) shared by the two variables [24]. As suggested in a number of studies [24, 28-30, 32, 34, 35, 37, 44], it is not always necessary to classify the model variables into endogenous or exogenous; therefore, following the notation in Drton et al. [35], the SEM representation of a cyclic graph can be given as follows

$$Y_i = \sum_{j \in Parent(i)} c_{ij} Y_j + \varepsilon_i, \qquad i, j = 1, \cdots, n, \tag{1}$$

where $c_{ij}$ denotes the weight of the directed edge $V_j \to V_i$, $\varepsilon_i$ denotes the random error that follows a certain distribution (Gaussian or non-Gaussian [31, 38]) with mean zero, and $Parent(i)$ denotes the set of parent nodes of node $i$. Without loss of generality, all $Y_i$s are assumed to be standardized via necessary transform [45]. To distinguish observed variables from latent variables, the superscripts *o* and *l* can be used (i.e., $Y_i^o$ and $Y_i^l$). Furthermore, let $\sigma_{ij} = \mathrm{Cov}(Y_i, Y_j)$ denote the covariance between two node variables. Also, let $\omega_{ij}$ denote the disturbance correlation between $Y_i$ and $Y_j$; by definition, $\omega_{ij} = 0$ if no undirected edge $V_j \leftrightarrow V_i$ can be found in $\mathbf{U}$. For convenience, we denote the covariance matrix and the disturbance correlation matrix as $\mathbf{\Sigma} = [\sigma_{ij}]$ and $\mathbf{\Omega} = [\omega_{ij}]$, respectively.

In general, the purpose of identifiability analysis is to verify whether certain unknown parameters can be uniquely and reliably determined for given model structures with or without considering data noise or model uncertainty [24, 28-30, 32, 34, 35, 37, 44]. Here the goal of structural identifiability analysis of SEMs is to determine whether the unknown parameters in matrices $\mathbf{C}$ and $\mathbf{\Omega}$ can be unambiguously determined for a given network structure $G=(\mathbf{V},\mathbf{D},\mathbf{U})$. This type of analysis does not take specific data distribution or noise level into consideration as its primary concern is not the robustness but the accuracy of parameter estimation via examining possible flaws in model structure or experimental design. More importantly, the structural identifiability of a parameter can be verified by checking its number of solutions to a system of polynomial equations. That is, a parameter is globally identifiable if only one solution exists, locally identifiable if a finite number of solutions exist, and unidentifiable if an infinite number of solutions exist [20].

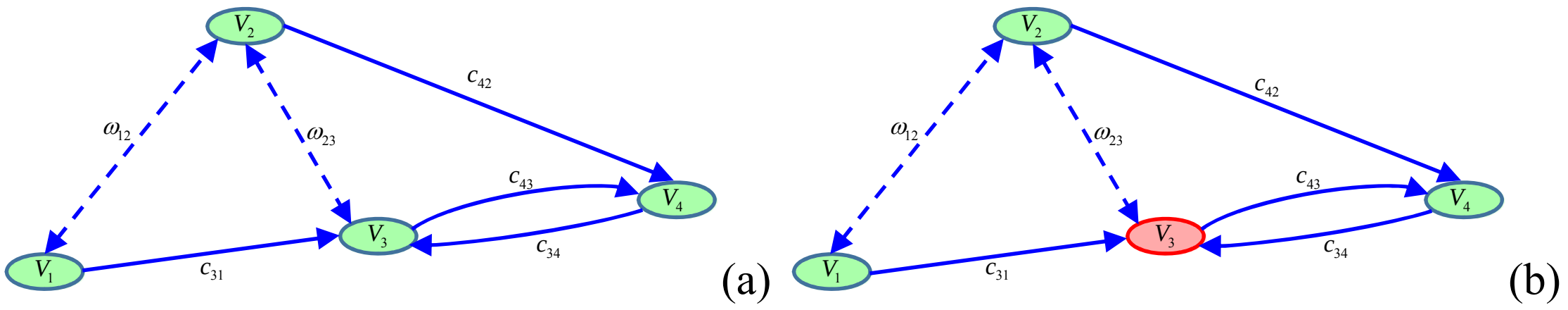

**Figure 1**. A mixed graph example, where the dashed bidirected arrow represents the disturbance correlation between the two variables. (a) Without latent variables; (b) With latent variable $Y_3$ (labelled in red).

For illustration purpose, we consider the mixed graph example in Fig. 1. The corresponding linear SEM is given as follows:

$$\begin{cases} Y_1 = \varepsilon_1 \\ Y_2 = \varepsilon_2 \\ Y_3 = c_{31}Y_1 + c_{34}Y_4 + \varepsilon_3 \\ Y_4 = c_{42}Y_2 + c_{43}Y_3 + \varepsilon_4 \\ \omega_{12} \neq 0 \\ \omega_{23} \neq 0 \end{cases}, \tag{2}$$

the coefficient and the disturbance correlation matrices are

$$\mathbf{C} = \begin{bmatrix} 0 & 0 & 0 & 0 \\ 0 & 0 & 0 & 0 \\ c_{31} & 0 & 0 & c_{34} \\ 0 & c_{42} & c_{43} & 0 \end{bmatrix} \text{ and } \mathbf{\Omega} = \begin{bmatrix} 0 & \omega_{12} & 0 & 0 \\ \omega_{12} & 0 & \omega_{23} & 0 \\ 0 & \omega_{23} & 0 & 0 \\ 0 & 0 & 0 & 0 \end{bmatrix}, \tag{3}$$

respectively, and the covariance matrices for Fig. 1(a) and 1(b) are

$$\mathbf{\Sigma}_a = \begin{bmatrix} \sigma_{11} & \sigma_{12} & \sigma_{13} & \sigma_{14} \\ \sigma_{12} & \sigma_{22} & \sigma_{23} & \sigma_{24} \\ \sigma_{13} & \sigma_{23} & \sigma_{33} & \sigma_{34} \\ \sigma_{14} & \sigma_{24} & \sigma_{34} & \sigma_{44} \end{bmatrix} \text{ and } \mathbf{\Sigma}_b = \begin{bmatrix} \sigma_{11} & \sigma_{12} & - & \sigma_{14} \\ \sigma_{12} & \sigma_{22} & - & \sigma_{24} \\ - & - & - & - \\ \sigma_{14} & \sigma_{24} & - & \sigma_{44} \end{bmatrix}, \quad (4)$$

respectively, where the symbol "$-$" denotes unknown covariance due to the existence of the latent variable $Y_3$. For the model corresponding to Fig. 1(b), the structure identifiability problem is to determine the number of solutions of each unknown parameter in matrices $\mathbf{C}$ and $\mathbf{\Omega}$ (i.e., $c_{31}$, $c_{34}$, $c_{42}$, $c_{43}$, $\omega_{12}$ and $\omega_{23}$).

**Generating identifiability equations**

Identifiability equations are obtained after eliminating all latent variables so they are a set of equations that contains only observed variables, unknown parameters and maybe other constants. It has been shown that under the assumption of normally-distributed disturbance, the covariance matrix $\mathbf{\Sigma}$ can be expressed in terms of $\mathbf{C}$ and $\mathbf{\Omega}$

$$\mathbf{\Sigma} = (\mathbf{I} - \mathbf{C})^{-\mathrm{T}} \mathbf{\Omega} (\mathbf{I} - \mathbf{C})^{-1}, \quad (5)$$

where $\mathbf{I}$ denotes the identity matrix. If the unknown covariance(s) in $\mathbf{\Sigma}$ can be eliminated, Eq. (5) will become a set of equations that involve only the unknown parameters in $\mathbf{C}$ and $\mathbf{\Omega}$ , and thus has been used as identifiability equations in previous studies [23, 34]. However, this approach needs to calculate the symbolic inversion of the matrix $(\mathbf{I} - \mathbf{C})$ such that it can only handle small models with a few unknown parameters even if with the use of the computer algebra tools [34]. Therefore, here we consider the Wright's method of path coefficients to generate identifiability equations [45, 46]. Briefly, the Wright's method considers the fact that two node variables are correlated with each other if there exists a path between these two nodes in a given network structure, and thus calculate the covariance between two node variables by adding the products of edge coefficients along each path. This approach can easily generate the identifiability equations in forms of nonlinear symbolic polynomials and has been previously verified and used for identifiability analysis of SEMs [29, 30].

More specifically, for an acyclic linear SEM (also called recursive SEM that corresponds to a directed acyclic graph), the covariance $\sigma_{ij}$ of a pair of variables $Y_i$ and $Y_j$ is calculated as $\sigma_{ij} = \sum_{path_k} \prod_{edge_l} \theta_l$, where $\theta_l$ is the coefficient of the $l$-th edge in Wright's path $k$ (i.e., $c_{pq}$ or $\omega_{pq}$ associated with a directed edge $V_q \to V_p$ or a bidirected edge $V_q \leftrightarrow V_p$). Note that each Wright's path must be unblocked [29, 30, 45, 46] (i.e., the consecutive sequence of edges of a Wright's path does not contain a pair of arrows that collide "head-to-head" in the corresponding graph). For a cyclic linear SEM (also called non-recursive), the directed graph part $\mathrm{G} = (\mathbf{V}, \mathbf{D})$ contains one or multiple cycles such that we need to enumerate all distinct cycles and paths. The key issue is that, for two nodes in the same cycle, there are two different sets of paths $V_i \to \cdots \to V_j$ and $V_j \to \cdots \to V_i$. Assume that the Wright's path coefficient method is also applicable to the paths on cycles. That is, two different sets of equations can be generated for $\sigma_{ij}$ and $\sigma_{ji}$, respectively, although $\sigma_{ij} = \sigma_{ji}$. Furthermore, for any latent variable $Y_i$ in a SEM, the covariance between $Y_i$ and any other variable is unknown and cannot be used to generate identifiability equations (see $\mathbf{\Sigma}_b$, the corresponding covariance matrix of Fig. 1(b)). In short, the existence of cycles or latent variables will lead to the increase or decrease of the number of identifiability equations, respectively, and thus will eventually affect the number of solutions of unknown model parameters.

Back to the examples in Fig. 1, it can be shown that the identifiability equations generated using the Wright's method for Fig. 1(a) and 1(b) are

$$\begin{cases} \sigma_{12} = \omega_{12} \\ \sigma_{13} = c_{31} + \omega_{12} c_{42} c_{34} \\ \sigma_{14} = c_{31} c_{43} + \omega_{12} c_{42} \\ \sigma_{23} = c_{42} c_{34} + \omega_{23} + \omega_{12} c_{31} \\ \sigma_{24} = c_{42} + \omega_{23} c_{43} + \omega_{12} c_{31} c_{43} \\ \sigma_{34} = c_{43} + \omega_{23} c_{42} + c_{31} \omega_{12} c_{42} \\ \sigma'_{34} = c_{34} + \omega_{23} c_{42} + c_{31} \omega_{12} c_{42} \end{cases}, \tag{6}$$

and

$$\begin{cases} \sigma_{12} = \omega_{12} \\ \sigma_{14} = c_{31} c_{43} + \omega_{12} c_{42} \\ \sigma_{24} = c_{42} + \omega_{23} c_{43} + \omega_{12} c_{31} c_{43} \end{cases}, \tag{7}$$

respectively. In Fig. 1(a), because the two nodes $V_3$ and $V_4$ are in the same cycle, we have $\sigma_{34} = c_{43} + \omega_{23}c_{42} + c_{31}\omega_{12}c_{42}$ for $V_3 \to V_4$ and $\sigma_{34}^{'} = c_{34} + \omega_{23}c_{42} + c_{31}\omega_{12}c_{42}$ for $V_4 \to V_3$ in Eq. (6) although $\sigma_{34}$ and $\sigma_{34}^{'}$ are the same. In Fig. 1(b), since the node $V_3$ is unobserved, the covariance $\sigma_{13}$, $\sigma_{23}$, $\sigma_{34}$ and $\sigma_{34}^{'}$ are unavailable for identifiability analysis as shown in Eq. (7).

**Generating identifiability matrices**

The identifiability equations are symbolic polynomials and are nonlinear with respect to unknown parameters. Simplifying and solving such equations using the computer algebra algorithms usually presents significant computational challenges [34]. Here we propose a novel and efficient approach, and the basic idea is to convert the identifiability equations to binary matrices, called identifiability matrices.

For each identifiability equations, one identifiability matrix is generated. More specifically, each column of the matrix corresponds to an unknown parameter, and each row corresponds to a monomial $\prod_{edge_l}\theta_l$. If the *i*-th monomial of an identifiability equation contains the *j*-th unknown parameter, then the corresponding matrix element $m_{ij}$ is equal to 1, otherwise $m_{ij} = 0$. Note that when generating the identifiability matrices, constant terms or known coefficients are not considered since they have no effects on the identifiability of unknown parameters. For illustration purpose, the list of identifiability matrices generated from Eq. (6) is given as follows

$$\begin{array}{c} \quad c_{31}\ c_{34}\ c_{42}\ c_{43}\ \omega_{12}\ \omega_{23} \end{array}$$

$$\sigma_{12}\begin{bmatrix}0 & 0 & 0 & 0 & 1 & 0\end{bmatrix},$$

$$\sigma_{13}\begin{bmatrix}1 & 0 & 0 & 0 & 0 & 0\\ 0 & 1 & 1 & 0 & 1 & 0\end{bmatrix},$$

$$\sigma_{14}\begin{bmatrix}1 & 0 & 0 & 1 & 0 & 0\\ 0 & 0 & 1 & 0 & 1 & 0\end{bmatrix},$$

$$\sigma_{23}\begin{bmatrix}0 & 1 & 1 & 0 & 0 & 0\\ 0 & 0 & 0 & 0 & 0 & 1\\ 1 & 0 & 0 & 0 & 1 & 0\end{bmatrix},$$

$$\sigma_{24}\begin{bmatrix}0&0&1&0&0&0\\0&0&0&1&0&1\\1&0&0&1&1&0\end{bmatrix},$$

$$\sigma_{34}\begin{bmatrix}0&0&0&1&0&0\\0&0&1&0&0&1\\1&0&1&0&1&0\end{bmatrix},$$

$$\sigma_{34}^{'}\begin{bmatrix}0&1&0&0&0&0\\0&0&1&0&0&1\\1&0&1&0&1&0\end{bmatrix}.$$

From Eq. (7), we can generate three matrices for $\sigma_{12}$, $\sigma_{14}$ and $\sigma_{24}$, respectively, which are the same as those from Eq. (6) and thus not shown here.

**Reducing identifiability matrices**

If all elements are 0 in an identifiability matrix $\mathbf{M}$, it is simply a zero matrix (denoted by $\mathbf{M}_Z$). Such matrices may occur during the reduction process. However, a zero matrix is not useful to identifiability analysis because it contains no unknown parameters. Therefore, once an identifiability matrix becomes a zero matrix after a certain number of reduction operations, it can be removed. For the same reason, a zero row in an identifiability matrix can also be deleted.

Given an identifiability matrix $\mathbf{M}$ with a row number $N_R(\mathbf{M})$ greater than 1, if all the rows in $\mathbf{M}$ are the same, such a matrix is called a repeated matrix (denoted by $\mathbf{M}_R$). The corresponding identifiability equation of a repeated matrix is $\sigma_{ij} = a_1\prod_l \theta_l + a_2\prod_l \theta_l \cdots + a_K\prod_l \theta_l$, where all the monomials are the same except for the constant coefficients $\{a_1, a_2, ..., a_K\}$ in the front. Since the equation can be simplified to $\sigma_{ij} = A\cdot\prod_l \theta_l$, where $A = a_1 + a_2 + \cdots + a_K$, the repeated identifiability matrix can be replaced by a single row without loss of information (denoted by $\mathbf{M}_{RI}$).

Further notations are needed to describe the relationships between two identifiability matrices. First, if all the rows in matrix $\mathbf{M}_2$ are from another matrix $\mathbf{M}_1$, $\mathbf{M}_2$ is called a sub-matrix of $\mathbf{M}_1$, denoted by $\mathbf{M}_2 = Sub(\mathbf{M}_1)$, and the remaining part is denoted by $Rem(\mathbf{M}_{1-2})$. Second, for two identifiability matrix $\mathbf{M}_1$ and $\mathbf{M}_2$ ($N_R(\mathbf{M}_1) \geq N_R(\mathbf{M}_2)$), if a sub-matrix of $\mathbf{M}_1$, denoted by $\mathbf{M}_3$, can be found such that it has the same number of rows as $\mathbf{M}_2$, and every element "1" in $\mathbf{M}_2$ is also a "1" in $\mathbf{M}_3$, then we call $\mathbf{M}_1$ includes $\mathbf{M}_2$, denoted by $\mathbf{M}_2 \subseteq \mathbf{M}_1$. An example of such a relationship is given in Fig. 2(a) for illustration purpose. Third, given two identifiability matrices $\mathbf{M}_1$ and $\mathbf{M}_2$ such that $N_R(\mathbf{M}_1) = N_R(\mathbf{M}_2)$ and $\mathbf{M}_2 \subseteq \mathbf{M}_1$, then $\mathbf{M}_3 = (\mathbf{M}_1 - \mathbf{M}_2)$ is called a complement matrix, denoted by $Comp(\mathbf{M}_1 - \mathbf{M}_2)$. See Fig. 2(b) for illustration of the complement matrix concept.

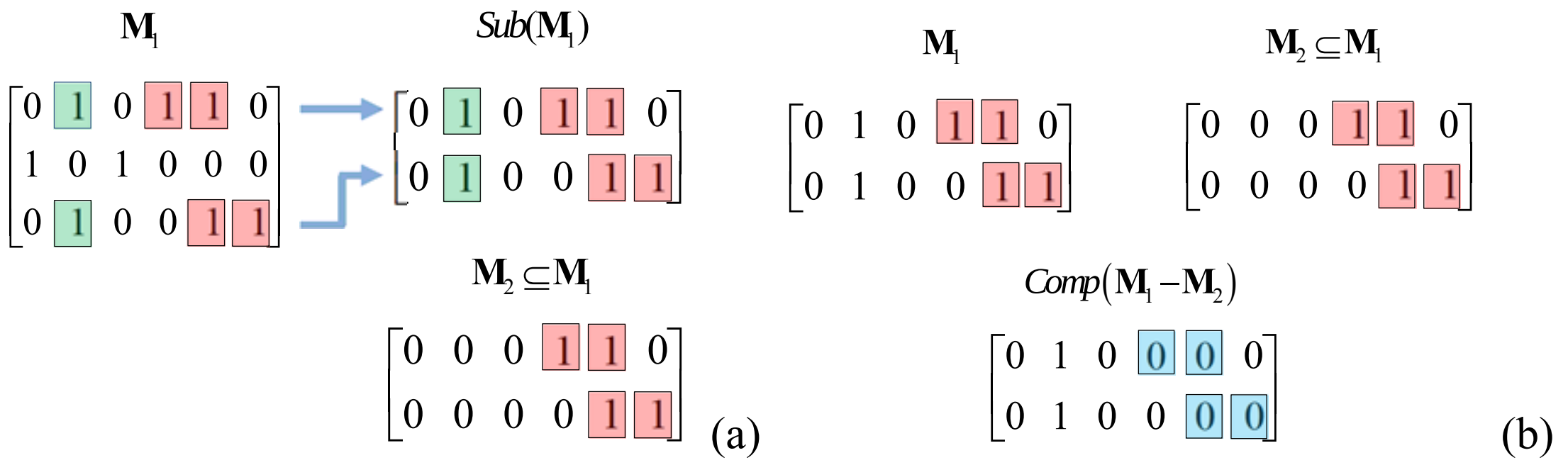

**Figure 2**. Illustration of (a) sub-matrix and matrix inclusion; and (b) complement matrix. Different colors are used to highlight the elements that remain the same or become different in different matrices.

Now the key issue is that the identifiability matrices before and after reduction should be equivalent; that is, the two sets of matrices should lead to the same conclusions on parameter identifiability. Let $\mathbf{M}_1 \sim \mathbf{M}_2$ denote two equivalent matrices, here we show that the following operations for matrix reduction can meet the requirement of identifiability equivalency (see Supplementary Material I for theoretical justification):

i) **Row swap**. Let $\boldsymbol{R}_i$ and $\boldsymbol{R}_j$ $(i \neq j)$ denote two different rows of an identifiability matrix $\mathbf{M}_1$, and let $\mathbf{M}_2$ denote the matrix generated after swapping $\boldsymbol{R}_i$ and $\boldsymbol{R}_j$, then $\mathbf{M}_1 \sim \mathbf{M}_2$.

ii) **Redundant row removal**. Let $\boldsymbol{R}_i$ and $\boldsymbol{R}_j$ $(i \neq j)$ denote two different rows of

an identifiability matrix $\mathbf{M}_1$. If $\boldsymbol{R}_i = \boldsymbol{R}_j$ and let $\mathbf{M}_2$ denote the matrix generated after removing $\boldsymbol{R}_i$ or $\boldsymbol{R}_j$, then $\mathbf{M}_1 \sim \mathbf{M}_2$.

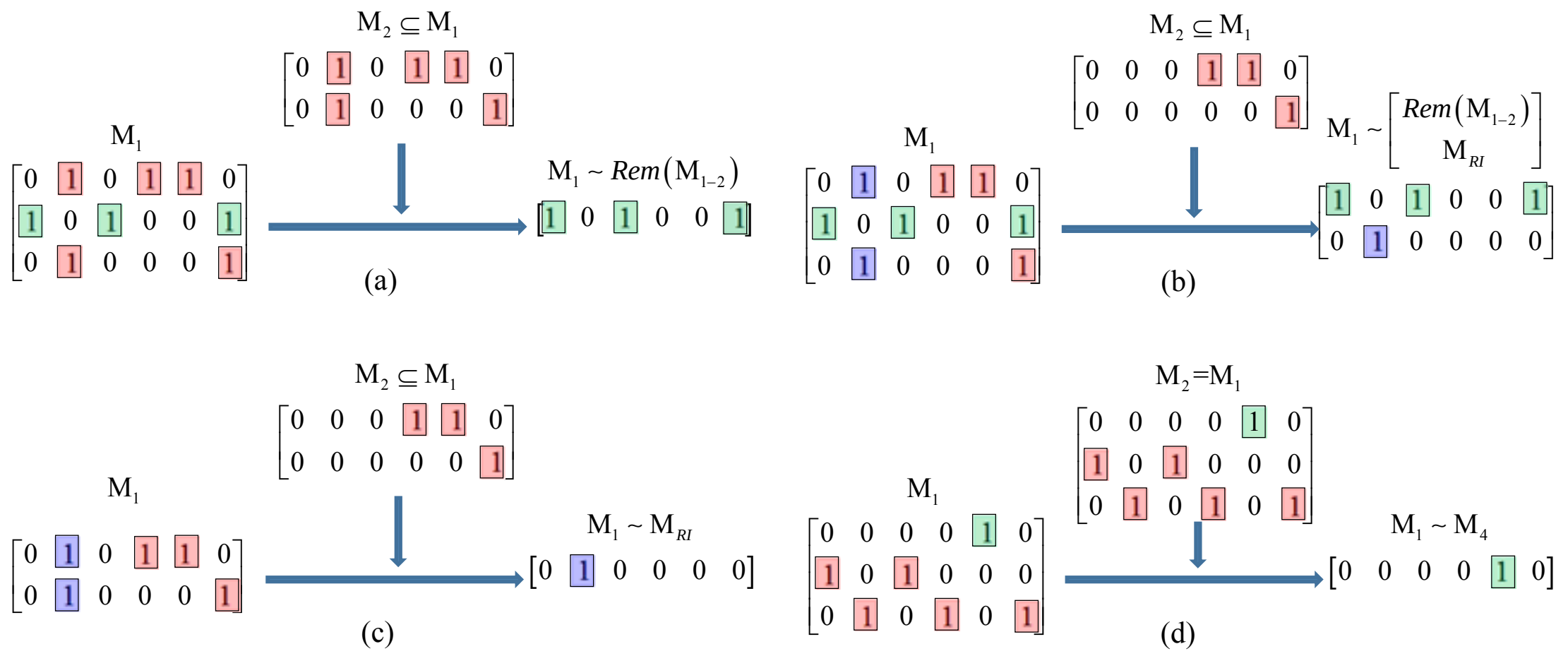

**Figure 3**. Several examples of the row deletion operation. Different colors are used to highlight the elements that remain the same or become different in different matrices.

iii) **Row deletion**. Let $\mathbf{M}_1$ and $\mathbf{M}_2$ be two identifiability matrices, which correspond to two different identifiability equations, such that $N_R(\mathbf{M}_1) > 1$ and $\mathbf{M}_2 \subseteq \mathbf{M}_1$. Also, let $\mathbf{M}_3 = sub(\mathbf{M}_1)$ be a sub-matrix consisting of $\mathbf{M}_1$'s rows that $\mathbf{M}_2$ has in $\mathbf{M}_1$. See Fig. 3 for examples.

- If $Rem(\mathbf{M}_{1-2}) \neq \mathbf{M}_Z$ and $Comp(\mathbf{M}_3 - \mathbf{M}_2) = \mathbf{M}_Z$, then $\mathbf{M}_1$ can be reduced to $Rem(\mathbf{M}_{1-2})$ without altering the parameter identifiability;
- If $Rem(\mathbf{M}_{1-2}) \neq \mathbf{M}_Z$ and $Comp(\mathbf{M}_3 - \mathbf{M}_2) = \mathbf{M}_R$, then $\mathbf{M}_1$ can be reduced to $\begin{bmatrix} Rem(\mathbf{M}_{1-2}) \\ M_{RI} \end{bmatrix}$ without altering the parameter identifiability;
- If $Rem(\mathbf{M}_{1-2}) = \mathbf{M}_Z$ and $Comp(\mathbf{M}_3 - \mathbf{M}_2) = \mathbf{M}_R$, then $\mathbf{M}_1$ can be reduced to $\mathbf{M}_{RI}$ without altering the parameter identifiability;
- If $Rem(\mathbf{M}_{1-2}) = \mathbf{M}_Z$ and $Comp(\mathbf{M}_3 - \mathbf{M}_2) = \mathbf{M}_z$ (i.e. $\mathbf{M}_1 = \mathbf{M}_2 = \mathbf{M}_3$), and take the row which has the least "1" elements in $\mathbf{M}_1$ to form a new matrix $\mathbf{M}_4$, then $\mathbf{M}_1$ can be reduced to $\mathbf{M}_4$ without altering the parameter identifiability.

The reduction process is iterative, and it stops until we cannot reduce the identifiability matrices further more. For illustration purpose, the reduction process for the identifiability matrices from Fig. 1(a) is shown in Fig. 4. The computation complexity of the reduction process depends on the number of rows in the identifiability matrices (denoted by $m$). In the worst scenario where every pair of rows need to be compared, the computing cost is $O(m^2)$; however, the efficiency can be improved if all the rows can be sorted before row comparison according to the positions of the "1" elements from left to right.

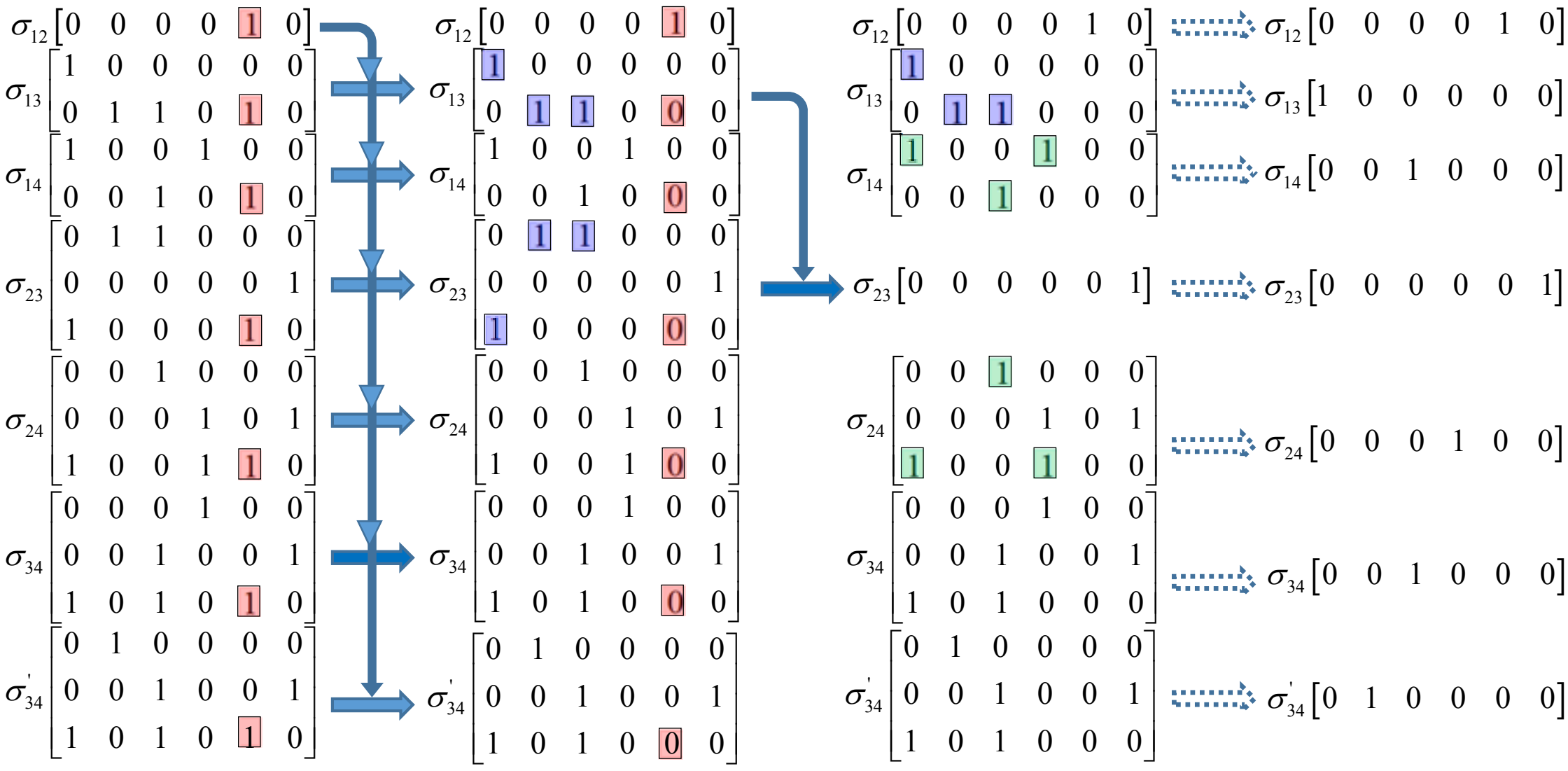

**Figure 4**. The reduction process of the identifiability matrices from Fig 1(a). In the left column, we subtract matrix $\sigma_{12}$ from matrix $\sigma_{13}$, $\sigma_{14}$, $\sigma_{23}$, $\sigma_{24}$, $\sigma_{34}$ and $\sigma_{34}'$; in the middle column, we then subtract matrix $\sigma_{13}$ from matrices $\sigma_{23}$; and so on; finally we get the reduction result in the right column.

### Determining parameter identifiability

After all identifiability matrices are reduced to the simplest forms using the operations described in the previous section, the identifiability of all the unknown parameters can be determined. The simplest case is to find out the globally identifiable. That is, if a matrix has only one row and this row has only one "1" element, the parameter corresponding to that "1" element is then globally identifiable because the associated identifiability equation is in the form $\theta_i = \text{const}$. For example, each matrix in the right column in Fig. 4 has only one row with only one "1" element, so each parameter corresponding to the "1" element is globally identifiable, i.e., the corresponding model of Fig. 1(a) is globally identifiable.

After removing all the matrices for globally identifiable parameters, the remaining matrices all have more than one "1" elements and they need to be regrouped and decoupled. That is, if the *i*-th columns of matrices $\mathbf{M}_1$ and $\mathbf{M}_2$ both contain one or more "1" elements, $\mathbf{M}_1$ and $\mathbf{M}_2$ will be in the same group. Here we describe the algorithm for grouping the identifiability matrices (see Fig. 5 for illustration).

(i) Apply the bit-OR operation to the first two rows, and then to the result and the 3rd row, and so on until the last row of a matrix to generate an indicator vector $\boldsymbol{R}_p$ such that each "1" element in this vector indicates the existence of a certain parameter;

(ii) Initialize an output vector $\boldsymbol{R}_{out}$ as the vector $\boldsymbol{R}_p$ that contains the largest number of "1" elements among all $\boldsymbol{R}_p$ s;

(iii) Check each of the $\boldsymbol{R}_p$ vectors to verify whether it has any common "1" element with $\boldsymbol{R}_{out}$ using the bit-AND operation. If the bit-AND result is not a zero vector, then the identifiability matrix corresponding to $\boldsymbol{R}_p$ will be added to the current group. Then update $\boldsymbol{R}_{out}$ by applying the bit-OR operation to $\boldsymbol{R}_{out}$ and the bit-AND result;

(iv) Repeat Step (iii) until no more matrices can be added to the current group;

(v) Remove all the matrices of the current group, and repeat steps (ii) to (iv) until all different groups are found.

The identifiability of all the parameters in the same group are determined together. According to the definition of identifiability matrix, one can tell that all the matrices of the same group correspond to a system of coupled polynomial equations, and the critical issue here is to determine the number of solutions of each parameter to these equations. Garcia and Li [47] have theoretically investigated this problem and shown that for a system of $n$ polynomial equations with $n$ complex variables, the number of solutions is equal to $q=\prod_{i=1}^{n} q_i$, where $q_i$ is the degree (the power of the highest ordered term) of equation $i$. Therefore, every unknown variable of the system has a unique solution when $q=1$, and has multiple solutions if $q>1$. Based on the work of Garcia and Li, we establish the theoretical connection between parameter identifiability and the grouped identifiability matrices, and the theoretical proof is given in Supplementary Material II for interested readers.

**Theorem 1** For the reduced identifiability matrices in the same group, let $N_M$ denote the number of matrices, let $N_P$ denote the number of unknown parameters, and let $N_{\max}$ be the maximum number of the "1" elements in one row of all the matrices.

- When $N_P > N_M$, all the parameters in the same group are unidentifiable;
- When $N_P = N_M$, the parameters are globally identifiable if $N_{\max} = 1$, and locally identifiable if $N_{\max} > 1$;
- When $N_P < N_M$, the parameters are at least locally identifiable.

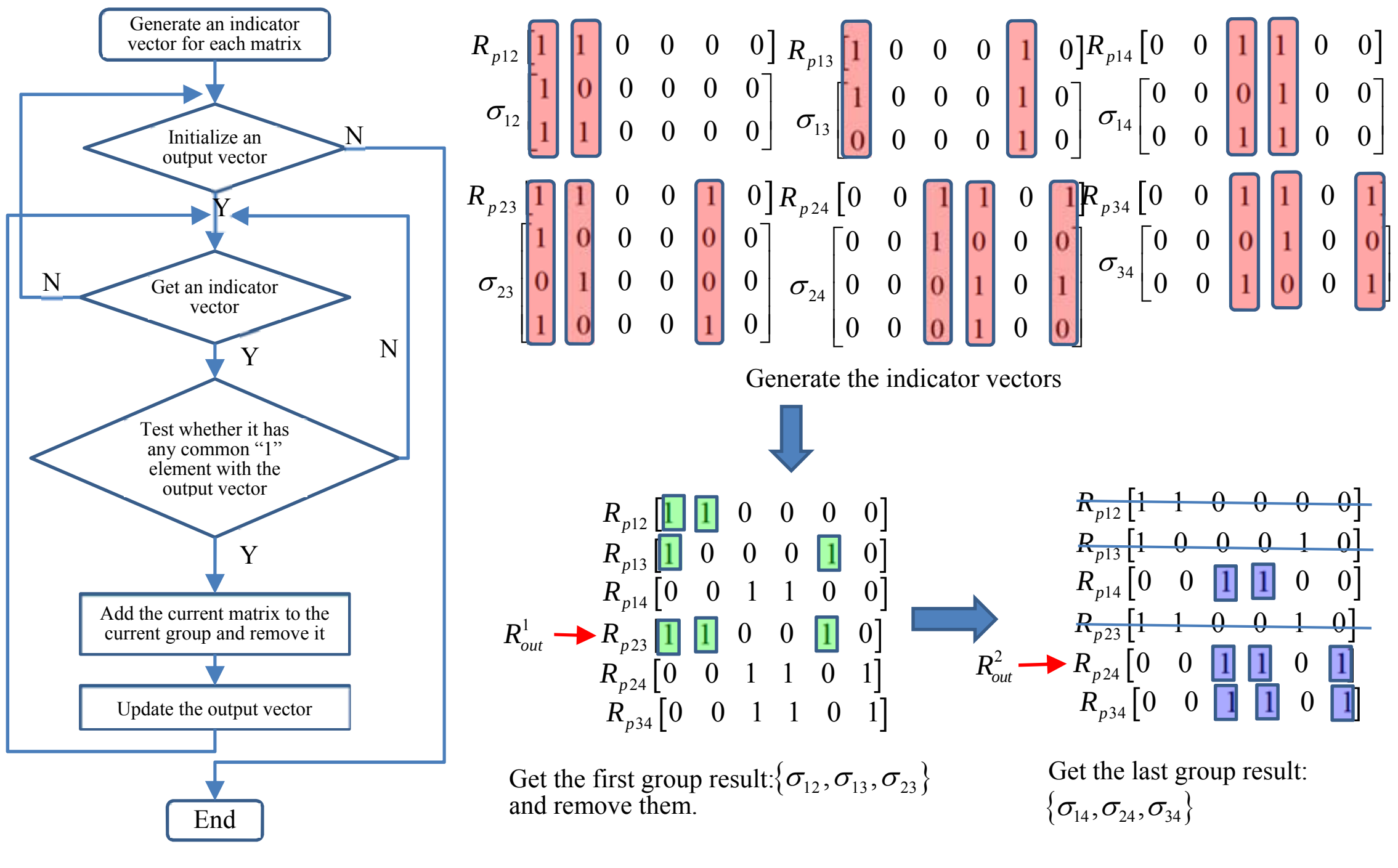

(a) The flowchart of reduction process (b) An example of the reduction process with two

**Figure 5**. Illustration of the grouping algorithm. (a) The reduction process flowchart; (b) An example of the reduction process with two matrix groups.

Based on Theorem 1, we can determine the structural identifiability of each parameter for the model in Fig. 1(b). As shown in Fig. 5(b), all the matrices are divided into two groups; and the number of matrices is $N_M$=3 , the number of unknown parameters is $N_P$=3, and $N_{\max}$=3 is greater than 1 in each group. Therefore, all the parameters corresponding to those "1" elements are locally identifiable. Similarly for the model in Fig. 1(b), one can tell $N_M$=3 and $N_P$=5 so all the parameters $\{c_{31}, c_{34}, c_{42}, c_{43}, \omega_{12}, \omega_{23}\}$ are unidentifiable.

# Results and discussion

## Overview of the framework

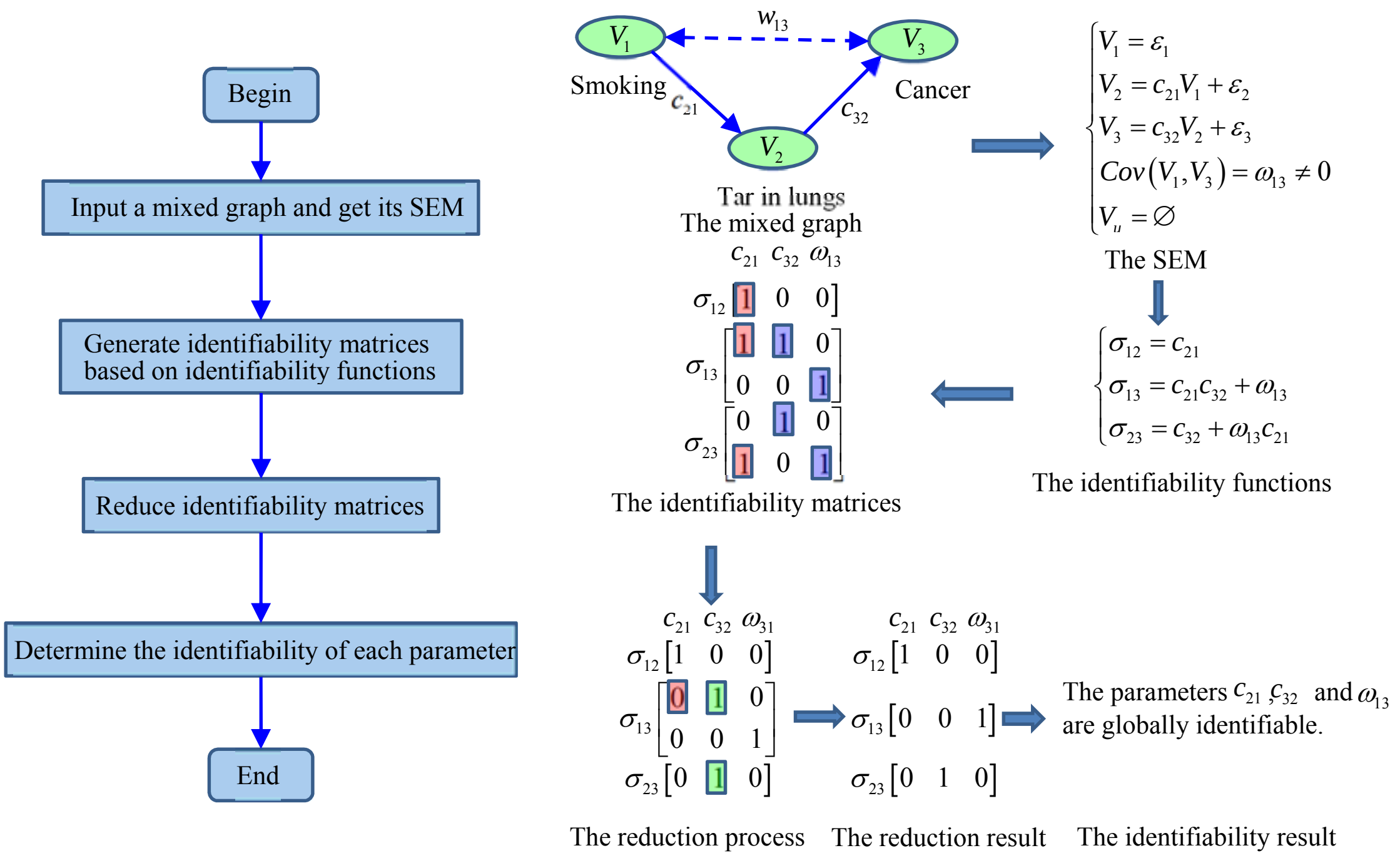

(a) Flow diagram

(b) The whole analysis process of the effect of smoking lung cancer

**Figure 6**. Illustration of the complete process of structural identifiability analysis: (a) Flowchart; (b) A simple example.

Graphical models have long been used to describe biological networks for a variety of important tasks like network structure identification. Many such quantitative analyses involve determination of unknown model parameters from experimental data, and identifiability analysis is a necessary step to perform before parameter estimation to assure the accuracy or robustness of the estimates. In particular, structural identifiability analysis can help to locate mis-specified substructures of models or improve experimental design with considering unobserved variables. A number of previous studies have proposed identifiability analysis techniques for structural equations models, with particular attention paid to specific network structures (e.g., directed acyclic graphs) or experimental conditions (e.g., without latent variables). Also, existing methods usually give an overall assessment instead of verifying the identifiability of each single parameter, and the use of symbolic computation algorithms (e.g., Gröbner basis reduction) is computationally expensive and has significantly limited the applications of these methods in more complex biological network structures and moderate to high-dimensional systems.

In this study, we develop a novel and efficient structural identifiability analysis technique to deal with a broader range of biological networks. To the best knowledge of the authors, the proposed method makes several worthwhile progresses in comparison with the previous work. First, the covariance between two observed variables can always be calculated (e.g., sample covariance) and thus treated as known, and a symbolic equation can be generated for this covariance by considering the effects of one variable on the other propagating through the path(s) between the two nodes. We adopt the Wright's path coefficient method [45, 46] for identifiability equation generation, which is not only more efficient than the approach of symbolic matrix inversion [34] but also can deal with cyclic networks with latent variables. Second, the computer algebra algorithms nowadays are only capable of efficiently solving nonlinear symbolic equations with a small number of variables, we propose a novel strategy to convert each symbolic equation to an identifiability matrix, and we also develop the necessary operations (e.g., row deletion) for identifiability matrix reduction without jeopardizing the equivalency of the identifiability results. Third, we present a strategy for regrouping the reduced identifiability matrices, and provide the guidelines with theoretical justification for determining parameter identifiability from the grouped matrices. The several contributions described above are in the same order of the algorithm pipeline, as depicted in Fig. 6(a). Finally, it should be stressed that the proposed algorithm is highly efficient because the main operations involved here are simple matrix manipulations like logical bitwise operations or matrix row deletion. For instance, it will take 0.3 to 4.5 seconds on a modern desktop computer to obtain the identifiability analysis results for a SEM with 4 nodes, 3 directed edges and 3 bidirected edges using the computer algebra method [34]; however, it will only take several milliseconds or less to reach the conclusions using the method proposed in this study as binary matrix operations are extremely efficient. It should be mentioned that many existing methods cannot be directly compared with the proposed method because they are not designed for static SEMs or they necessarily require human intervention. For example, DAISY has been proposed for determining parameter identifiability of ODE models [48]; and the method of identifiability tableaus [49] is based on Jacobian matrix that involves partial derivatives, while our method is based on a system of polynomial equations and does not require the calculation of derivatives.

**Verification using benchmark models**

In order to verify the validity of the proposed method, we have collected a number of benchmark models available in public literature to check whether the identifiability results obtained using our method are consistent with those obtained by other existing methods. Since these existing models do not contain any latent variable, we also consider a model with latent variables at the end of this section to show the capacity of our method.

The first benchmark model is for investigating the effects of smoking on lung cancer [24], the graph contains three nodes (variables), two directed edges, and one bidirected edge (disturbance correlation). All the parameters in this model are found to be globally identifiable and the detailed analysis process have been shown in Fig. 6(b). The second benchmark model was previously studied by Sullivant et al. [34], and its graph contains three nodes, one directed edge, and two bidirected edges. Again, all the parameters in the second model turn out to be globally identifiable and the analysis details are given in Supporting Materials III. The third benchmark model investigated by Drton et al. [35] is for an acyclic graph with four nodes, three directed edges, and three bidirected edges. From the same literature (Ref. [35]), we collected the fourth benchmark model that is more complicated in terms of number of variables and their interactions. The fifth benchmark model derived from the work of Kline el al. [22] is a cyclic graph with six nodes, six directed edges and three bidirected edges. The purpose of this model is to show that the proposed approach can deal with cyclic graphs. We derived the sixth benchmark model from the work of Drton et al. [35]. This cyclic graph has six nodes, six directed edges, and three bidirected edges; however, for this model, we also considered the case of multigraph (i.e., there exist both a directed edge and an bidirected edge between two nodes), which has been paid particular attention in the previous study of Brito and Pearl [36]. We reported the structural identifiability analysis details and results of the third to sixth models also in Supporting Materials III.

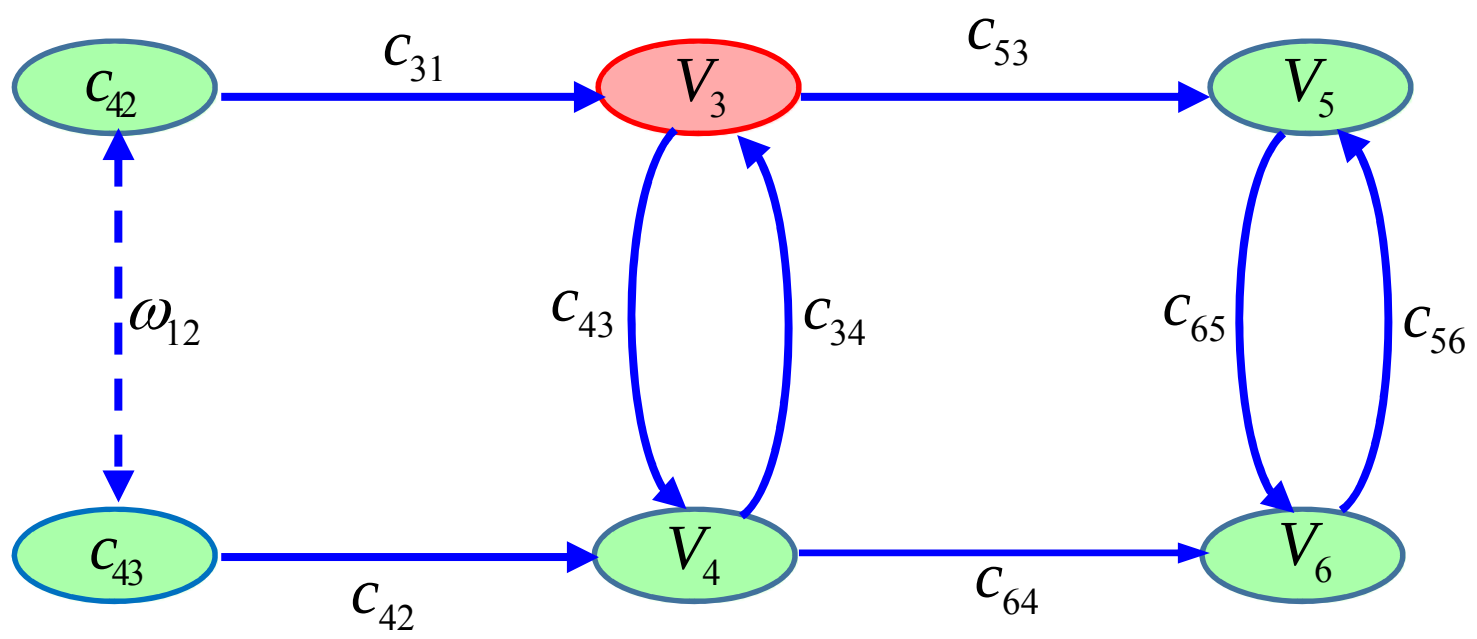

**Figure 7**. A mixed graph with feedback loops and one latent variable.

While the identifiability results obtained using our method for all the benchmark models above are consistent with the conclusions in the existing literature, we have not found a model with explicit latent variables in literature. We thus derived such a model from the work of Kline [23] by assuming that node $V_3$ is unobserved. As shown in Fig. 7, the mixed graph has 6 nodes, 2 cycles, and one latent variable $V_3$ (labelled in red). There are 9 parameters $\{c_{31}, c_{34}, c_{42}, c_{43}, c_{53}, c_{56}, c_{64}, c_{65}, \omega_{12}\}$ in this model. Because the latent variable $Y_3$ is not observed, the covariance between $Y_3$ and other variables is unavailable for identifiability analysis. Therefore, only the following identifiability equations can be generated:

$$\begin{cases} \sigma_{12} = \omega_{12} \\ \sigma_{14} = c_{31}c_{43} + \omega_{12}c_{42} \\ \sigma_{15} = c_{31}c_{53} + c_{31}c_{43}c_{64}c_{56} + \omega_{12}c_{42}c_{34}c_{53} + \omega_{12}c_{42}c_{64}c_{56} \\ \sigma_{16} = c_{31}c_{53}c_{65} + c_{31}c_{43}c_{64} + \omega_{12}c_{42}c_{64} + \omega_{12}c_{42}c_{34}c_{53}c_{65} \\ \sigma_{24} = c_{42} + \omega_{12}c_{31}c_{43} \\ \sigma_{25} = c_{42}c_{64}c_{56} + c_{42}c_{34}c_{53} + \omega_{12}c_{31}c_{53} + \omega_{12}c_{31}c_{43}c_{64}c_{56} \\ \sigma_{26} = c_{42}c_{64} + c_{42}c_{34}c_{53}c_{65} + \omega_{12}c_{31}c_{53}c_{65} + \omega_{12}c_{31}c_{43}c_{64} \\ \sigma_{45} = c_{34}c_{53} + c_{64}c_{56} + c_{42}\omega_{12}c_{31}c_{53} \\ \sigma_{46} = c_{64} + c_{34}c_{53}c_{65} + c_{42}\omega_{12}c_{31}c_{53}c_{65} \\ \sigma_{56} = c_{56} + c_{64}c_{34}c_{53} + c_{64}c_{42}\omega_{12}c_{31}c_{53} \\ \sigma'_{56} = c_{65} + c_{53}c_{43}c_{64} + c_{64}c_{42}\omega_{12}c_{31}c_{53} \end{cases} . \quad (8)$$

The identifiability matrices in Fig. 8(a) can be generated according to the identifiability equations above, and these identifiability matrices are then reduced following the process shown in Fig. 8(b). Finally, the reduction results in Fig. 8(c) are obtained, from which we can tell that the matrices associated with $\sigma_{16}$ and $\sigma_{46}$ are the same, and the matrices associated with $\sigma_{14}$ and $\sigma'_{56}$ are also the same. This observation suggests that there exist two redundant identifiability equations. Also, one can tell from Fig. 8(c) that all the matrices have only one row with one “1” element. Therefore, all model parameters are globally identifiable despite the existence of a latent variable. This example model thus illustrates the capability of the proposed approach handling models with latent variables.

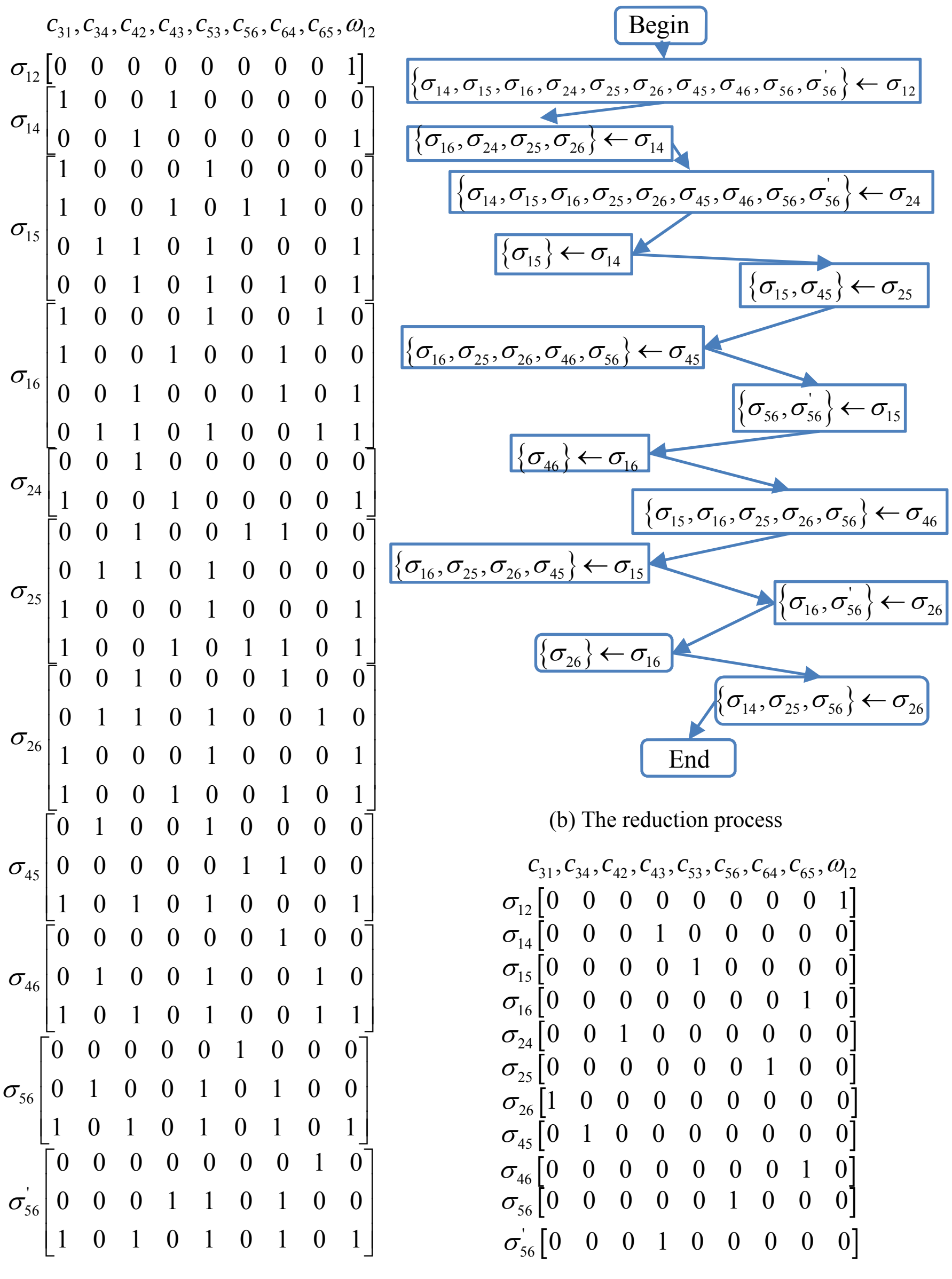

(a) The original identifiability matrices (b) The reduction process (c) The reduction result

**Figure 8**. The reduction process of Fig. 7. The left arrow means that the left matrices are reduced by the right matrices. (a) The original identifiability matrices; (b) The reduction process; (c) The reduction results.

## Applications to real biological networks

Numerous biological networks can be found in a variety of databases or knowledge repositories [50, 51]; limited by resources, here we only consider a subnetwork structure of the within-host influenza virus life cycle as an application example. More specifically, influenza A virus (IAV) can infect multiple species including birds and human, and it has long been a major threat to public health by causing seasonal epidemics or sporadic pandemics [52]. A systematic understanding of IAV infection and immune response mechanisms is thus of significant scientific interest nowadays. For this purpose, a comprehensive map of the influenza virus life cycle together with molecular-level host responses has been previously constructed from hundreds of related publications by Matsuoka et al. [53], including several critical network modules like virus entry, virus replication and transcription, post-translational processing, transportation of virus proteins, and packaging and budding. Here we choose the subnetwork of virus replication, to which particular attention has been paid by many previous experimental studies [54-57].

However, influenza A virus replication is a complex process, involving many different biomolecules. It is therefore usually infeasible for one single experimental study to observe all the components and their interactions simultaneously, leading to the presence of latent variables. In addition, such complex molecular interactions cannot always be described by a directed acyclic graph due to the existence of, e.g., feedback loops. Therefore, we consider the IAV replication network module as a suitable example of cyclic graphical models with latent variables. We thus derived the mixed graph in Fig. 9(a) from Matsuoka's work [53], which contains 22 nodes, 30 edges, and one cycle. The 5 pre-selected latent variables are labelled in red, and the observed nodes are in green. After applying the proposed algorithm to this network structure, the structure identifiability analysis result is visualized in Fig. 9(b), where 16 globally identifiable edge coefficients are in green, 6 locally identifiable edge coefficients in blue, and 8 unidentifiable edge coefficients in red. From the results in Fig. 9(b), we can also tell that local network topological structures may have an important effect on parameter identifiability. For example, the NP inhibitor node has an in-degree 0 and is unobserved, which is the direct reason why all the edges starting from such a node are unidentifiable. In addition, both the cRNA and cRNP nodes have a comparatively high total degree (an in-degree 4 and an out-degree 1 for both nodes); however, the cRNP node is unobserved such that all the edges connected with it are unidentifiable, while the four incoming edges to the cRNA nodes are globally identifiable. The implication of such observations on experimental design is that, the nodes with an in-degree or out-degree 0 and the nodes with a high total degree (e.g., hub genes) are suggested to be experimentally observed to reduce the identifiability problem.

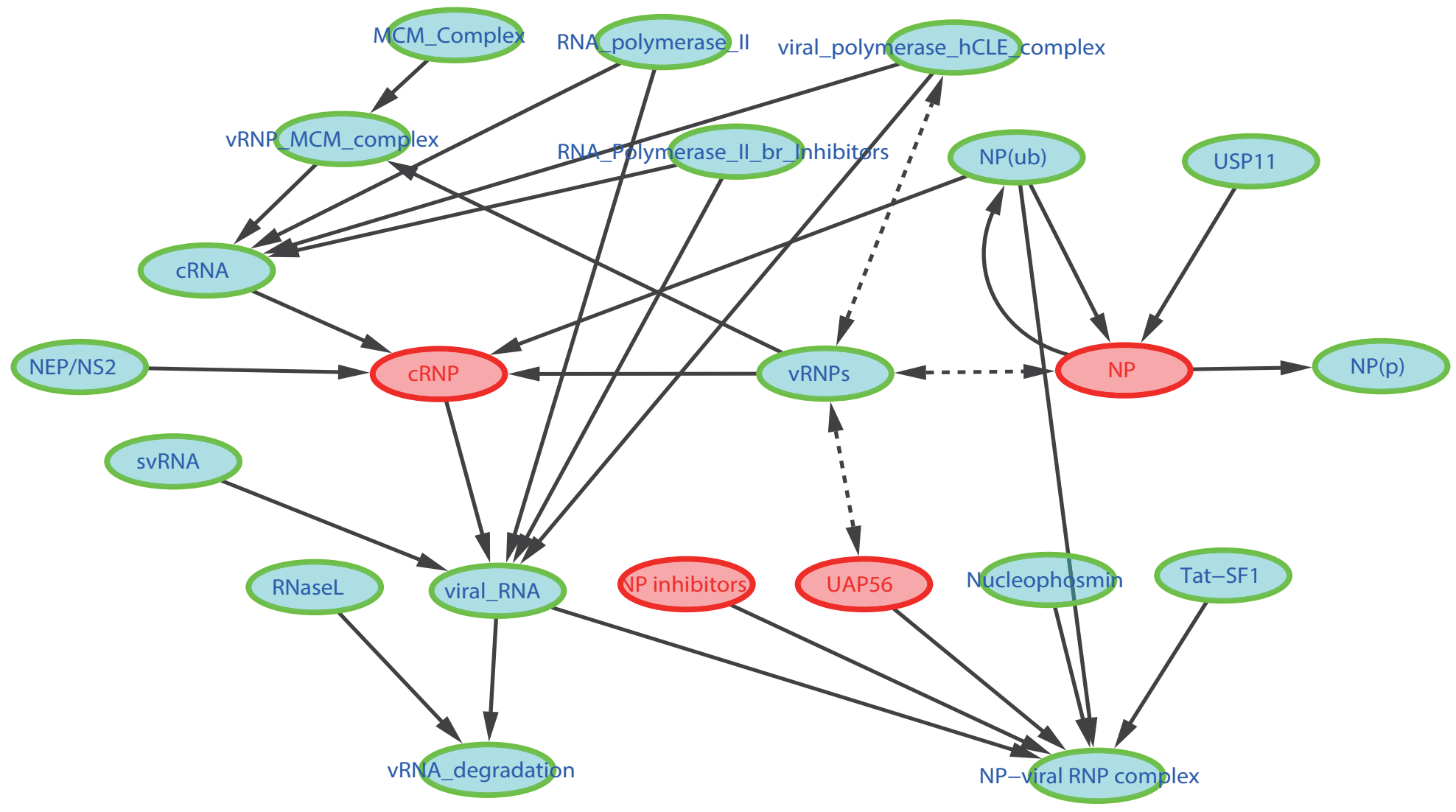

(a) The mixed graph.

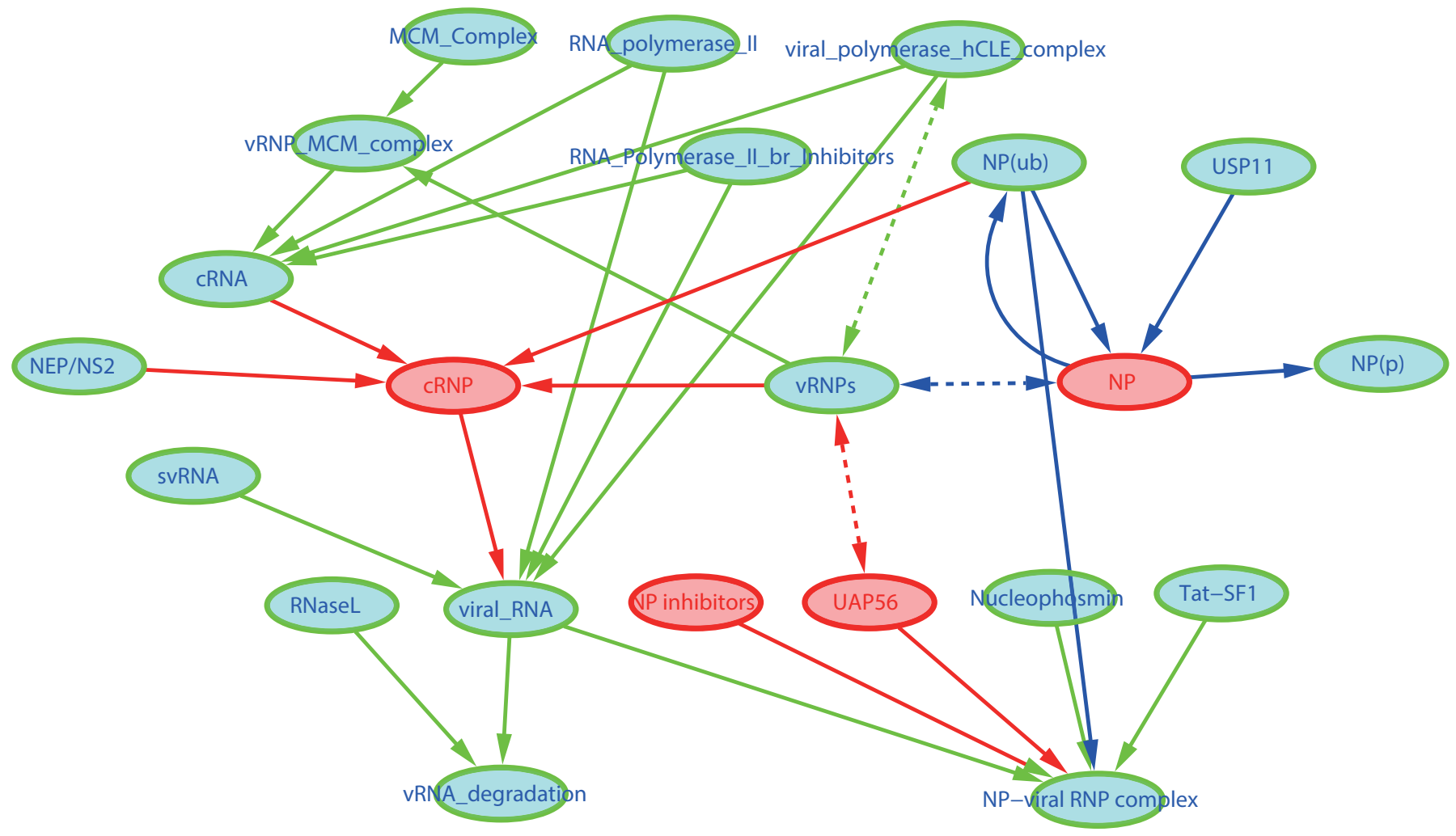

(b) The analysis result.

**Figure 9**. Identifiability analysis of the influenza A virus replication module. The read nodes are unobserved variables and the green nodes are observed variables in both (a) and (b). In (b), the globally identifiable edge coefficients are in green, the locally identifiable coefficients are in blue, and the unidentifiable coefficients are in red.

# Conclusions

In this study, we proposed a novel method for structural identifiability analysis of cyclic graphical models with explicit latent variables. Briefly, to deal with a broader range of network structures, the Wright's path coefficient method is adapted to generate the identifiability equations and particular attention has been paid to cyclic mixed graphs (as well as the multigraph case, see Benchmark Model 5 in Supplementary Material III) with explicit latent variables. To achieve high computing efficiency, the identifiability equations are converted to binary identifiability matrices and the necessary strategies have been developed for matrix reduction and regrouping. Parameter identifiability can then be verified at the single parameter level based on the reduced and grouped identifiability matrices after a connection between the number of non-zero matrix elements and the theoretical work of Garcia and Li. The validity of the proposed approach was theoretically justified and further verified using existing benchmark models. In addition, the proposed approach was applied to a real network structure for influenza A virus replication to gain insights into experimental design.

In summary, this study provides a basis for efficient model refinement and informative experiment design, and thus may facilitate investigators to expedite our understanding of network structure and interaction mechanisms in complex biological systems. However, we recognize that many real biological networks are high-dimensional with complex nonlinear interactions. Therefore, the proposed approach will need to be extended to deal with more realistic problems in the future.

## Availability of data and material

The network structure data used in this study are all selected from public literature, including the FluMap database [53].

## Funding

This work was partially supported by the Fundamental Research Funds for the Central Universities of China (ZYGX2014J064).

## Supporting Materials I: Identifiability Preservation by Matrix Reduction

In this study, the identifiability equations of a SEM are generated by Wright's path coefficient method. That is, the covariance $\sigma_{ij}$ between a pair of variables $V_i$ and $V_j$ is equal to $\sum_{path_k}\prod_{edge_l}\theta_l$. Note that each monomial $\prod_{edge_l}\theta_l$ corresponds to an non-redundant path between $V_i$ and $V_j$, thus there exist no identical monomials $\prod_{edge_l}\theta_l$ in all the identifiability equations. This observation also suggests that none of the identifiability equations can be expressed in terms of linear combinations of other equations, which is called the non-redundancy property.

Before we show the identifiability preservation by identifiability matrix reduction, the following definitions need to be introduced.

**Definition 1 (Equivalent Identifiability Equations)** If two identifiability equations $f_1(\theta)$ and $f_2(\theta)$ can be reduced to the same Gröbner basis, then $f_1(\theta)$ and $f_2(\theta)$ are called equivalent, denoted by $f_1(\theta) \sim f_2(\theta)$. □

**Definition 2 (Equivalent Identifiability Matrices)** For two identifiability matrices $\mathbf{M}_1$ and $\mathbf{M}_2$, if the two corresponding identifiability equations are equivalent, then $\mathbf{M}_1$ and $\mathbf{M}_2$ are also equivalent, denoted by $\mathbf{M}_1 \sim \mathbf{M}_2$. □

**Remark 1**. According to Definition 1, it is straightforward to tell that addition or multiplication of a constant to a monomial term will produce an equivalent equation. For example, given $f_1(\omega,c)$ : $\sigma_{12} = a_1\omega_{12} + a_2c_{31}\omega_{23}$ , $f_2(\omega,c)$ : $\sigma_{12} - 5 = a_1\omega_{12} + a_2c_{31}\omega_{23}$ , $f_3(\omega,c)$ : $\sigma_{12} = 3a_1\omega_{12} + a_2c_{31}\omega_{23}$ and $f_4(\omega,c)$ : $\sigma_{12} - 2 = 3a_1\omega_{12} + 4a_2c_{31}\omega_{23}$ , where $a_1$ and $a_2$ are the constant coefficients of monomials, then $f_1(\omega,c) \sim f_2(\omega,c) \sim f_3(\omega,c) \sim f_4(\omega,c)$. □

According to Definition 2, we can introduce three categories of operations on identifiability matrices, which will preserve the identifiability of the original system.

i) **Row swap**. Let $\boldsymbol{R}_i$ and $\boldsymbol{R}_j$ $(i \neq j)$ denote two different rows of an identifiability matrix $\mathbf{M}_1$, and let $\mathbf{M}_2$ denote the matrix generated after swapping $\boldsymbol{R}_i$ and $\boldsymbol{R}_j$ ,

then $\mathbf{M}_1 \sim \mathbf{M}_2$.

**Proof**. According to the generation rule of identifiability matrices, the row $\boldsymbol{R}_i$ represents the *i*-th monomial $\prod_{edge_l} \theta_l^{(i)}$ of an identifiability equation $\sigma = \sum_{path_k} \prod_{edge_l} \theta_l$ ; similarly, the row $\boldsymbol{R}_j$ represents the *j*-th monomial $\prod_{edge_l} \theta_l^{(j)}$ of the same identifiability equation. Swapping two rows $\boldsymbol{R}_i \leftrightarrow \boldsymbol{R}_j$ is equivalent to swap the positions of the two monomials in the identifiability equation, which will not change the identifiability equation according to the communitive law $\prod_{edge_l} \theta_l^{(i)} + \prod_{edge_l} \theta_l^{(j)} = \prod_{edge_l} \theta_l^{(j)} + \prod_{edge_l} \theta_l^{(i)}$. Therefore, $\mathbf{M}_1$ is equivalent to $\mathbf{M}_2$. ■

ii) **Redundant row removal**. Let $\boldsymbol{R}_i$ and $\boldsymbol{R}_j$ $(i \neq j)$ denote two different rows of an identifiability matrix $\mathbf{M}_1$. If $\boldsymbol{R}_i = \boldsymbol{R}_j$ and let $\mathbf{M}_2$ denote the matrix generated after removing $\boldsymbol{R}_i$ or $\boldsymbol{R}_j$ , then $\mathbf{M}_1 \sim \mathbf{M}_2$.

**Proof.** If $\boldsymbol{R}_i = \boldsymbol{R}_j$, the corresponding monomials are the same (maybe expect for the constant coefficients in front). The two monomials can thus be merged into one monomial term, which indicates that $\mathbf{M}_2$ is equivalent to $\mathbf{M}_1$. ■

iii) **Row deletion**. Let $\mathbf{M}_1$ and $\mathbf{M}_2$ be two identifiability matrices, which correspond to two different identifiability equations, such that $N_R(\mathbf{M}_1) > 1$ and $\mathbf{M}_2 \subseteq \mathbf{M}_1$. Also, let $\mathbf{M}_3 = sub(\mathbf{M}_1)$ be a submatrix consisting of $\mathbf{M}_1$'s rows that $\mathbf{M}_2$ has in $\mathbf{M}_1$. See Fig. 3 for examples.

- If $Rem(\mathbf{M}_{1-2}) \neq \mathbf{M}_Z$ and $Comp(\mathbf{M}_3 - \mathbf{M}_2) = \mathbf{M}_Z$, then $\mathbf{M}_1$ can be reduced to $Rem(\mathbf{M}_{1-2})$ without altering the parameter identifiability;
- If $Rem(\mathbf{M}_{1-2}) \neq \mathbf{M}_Z$ and $Comp(\mathbf{M}_3 - \mathbf{M}_2) = \mathbf{M}_R$, then $\mathbf{M}_1$ can be reduced

to $\begin{bmatrix} Rem(\mathbf{M}_{1-2}) \\ M_{RI} \end{bmatrix}$ without altering the parameter identifiability;

- If $Rem(\mathbf{M}_{1-2}) = \mathbf{M}_Z$ and $Comp(\mathbf{M}_3 - \mathbf{M}_2) = \mathbf{M}_R$, then $\mathbf{M}_1$ can be reduced to $\mathbf{M}_{RI}$ without altering the parameter identifiability;
- If $Rem(\mathbf{M}_{1-2}) = \mathbf{M}_Z$ and $Comp(\mathbf{M}_3 - \mathbf{M}_2) = \mathbf{M}_z$ (i.e. $\mathbf{M}_1 = \mathbf{M}_2 = \mathbf{M}_3$), and take the row which has the least "1" elements in $\mathbf{M}_1$ to form a new matrix $\mathbf{M}_4$, then $\mathbf{M}_1$ can be reduced to $\mathbf{M}_4$ without altering the parameter identifiability.

**Proof.** (1) If $Rem(\mathbf{M}_{1-2}) \neq \mathbf{M}_Z$ and $Comp(\mathbf{M}_3 - \mathbf{M}_2) = \mathbf{M}_Z$, we know that $\mathbf{M}_1$ consists of $\mathbf{M}_2$ and $Rem(\mathbf{M}_{1-2})$. That is, the identifiability equation corresponding to $\mathbf{M}_1$ can be divided into two parts. The part corresponding to $\mathbf{M}_2$ can be denoted as $expr(\mathbf{M}_2)$, and the other part corresponding to $Rem(\mathbf{M}_{1-2})$ can be denoted as $expr(Rem(\mathbf{M}_{1-2}))$. Let $\sigma_1$ and $\sigma_2$ denote the covariance corresponding to $\mathbf{M}_1$ and $\mathbf{M}_2$, respectively, then $\sigma_1 = expr(\mathbf{M}_2) + expr(Rem(\mathbf{M}_{1-2}))$ and $\sigma_2 = expr(\mathbf{M}_2)$. Using simple algebraic operations, we obtain $\sigma_1 - \sigma_2 = expr(Rem(\mathbf{M}_{1-2}))$. Since $\sigma_2$ is a known constant, according to Remark 1, we know that $\mathbf{M}_1$ can be reduced to $Rem(\mathbf{M}_{1-2})$ without altering the parameter identifiability.

(2) Let $\sigma_1$ and $\sigma_2$ be the known covariance corresponding to $\mathbf{M}_1$ and $\mathbf{M}_2$, respectively, that is, $\sigma_1 = expr(M_1)$ and $\sigma_2 = expr(M_2)$. If $Rem(\mathbf{M}_{1-2}) \neq \mathbf{M}_Z$ and $Comp(\mathbf{M}_3 - \mathbf{M}_2) = \mathbf{M}_R$, then certain monomials in the identifiability equation corresponding to $\mathbf{M}_1$ will have a common term such that $\sigma_1 = expr(Rem(\mathbf{M}_{1-2})) + expr(\mathbf{M}_{RI}) \times expr(\mathbf{M}_2)$. Replace $expr(\mathbf{M}_2)$ with $\sigma_2$ and obtain $\sigma_1 = expr(Rem(\mathbf{M}_{1-2})) + \sigma_2 expr(\mathbf{M}_{RI})$. It is known from Lemma 1 that changing the coefficient of a monomial does not change its identifiability matrix of an identifiability equation. So $\mathbf{M}_1$ can be reduced to $\begin{bmatrix} Rem(\mathbf{M}_{1-2}) \\ M_{RI} \end{bmatrix}$ without altering the parameter identifiability.

(3) Let $\sigma_1$ and $\sigma_2$ be the known covariance corresponding to $\mathbf{M}_1$ and $\mathbf{M}_2$, respectively, that is, $\sigma_1 = expr(\mathbf{M}_1)$ and $\sigma_2 = expr(\mathbf{M}_2)$. If $Rem(\mathbf{M}_{1-2}) = \mathbf{M}_Z$ and $Comp(\mathbf{M}_3 - \mathbf{M}_2) = \mathbf{M}_R$, then all the monomials in the identifiability equation corresponding to $\mathbf{M}_1$ share a common term such that $\sigma_1 = expr(\mathbf{M}_{RI}) \times expr(\mathbf{M}_2)$. Replace $expr(\mathbf{M}_2)$ by $\sigma_2$ to obtain $\sigma_1 = \sigma_2 expr(\mathbf{M}_{RI})$. According to Lemma 1, changing the coefficient of a monomial does not change its identifiability matrix in. Therefore, $\mathbf{M}_1$ can be reduced to $\mathbf{M}_{RI}$ without altering the parameter identifiability.

(4) If $Rem(\mathbf{M}_{1-2}) = \mathbf{M}_Z$ and $Comp(\mathbf{M}_3 - \mathbf{M}_2) = \mathbf{M}_z$, we have $\mathbf{M}_1 = \mathbf{M}_2 = \mathbf{M}_3$. The matrix $\mathbf{M}_4$ has only one row, which is the row with the least number of ones in $\mathbf{M}_1$, so we can express $\mathbf{M}_1$ as $\mathbf{M}_1 = \begin{bmatrix} \mathbf{M}_4 \\ Rem(\mathbf{M}_{1-4}) \end{bmatrix}$. Now let $\sigma_1$ and $\sigma_2$ be the known covariance corresponding to $\mathbf{M}_1$ and $\mathbf{M}_2$, respectively, then we have

$$\sigma_1 = a_1 \cdot expr(\mathbf{M}_4) + a_2 \cdot expr(Rem(\mathbf{M}_{1-4})),$$

$$\sigma_2 = b_1 \cdot expr(\mathbf{M}_4) + b_2 \cdot expr(Rem(\mathbf{M}_{1-4})),$$

where $a_1$, $a_2$, $b_1$ and $b_2$ are the nonzero constant coefficients. It is known from the non-redundant property that these two equations are linearly independent. So from the latter equation we have $expr(Rem(\mathbf{M}_{1-4})) = \sigma_2 / b_2 - b_1 / b_2 \times expr(\mathbf{M}_4)$. Replace $expr(Rem(\mathbf{M}_{1\text{-}4}))$ by $\sigma_2 / b_2 - b_1 / b_2 \times expr(\mathbf{M}_4)$ in the equation for $\sigma_1$ to obtain

$$\sigma_1 = a_1 \times expr(\mathbf{M}_4) + a_2\sigma_2 / b_2 - a_2 b_1 / b_2 \times expr(\mathbf{M}_4). \quad \text{(I.1)}$$

Thus, we can rewrite this equation as

$$\sigma_3 = a_3 \times expr(\mathbf{M}_4), \quad \text{(I.2)}$$

where $\sigma_3 = \sigma_1 - a_2\sigma_2 / b_1$, $a_3 = a_1 - a_2 b_1 / b_2$ and $a_3 \neq 0$. Since (I.1) and (I.2) have the same Gröbner basis, we know $\mathbf{M}_1$ can be reduced to $\mathbf{M}_4$ without altering the parameter identifiability. ■

## Supporting Materials II —— Proof of Theorem 1

**Theorem 1** For the reduced identifiability matrices in the same group, let $N_M$ denote the number of matrices, let $N_P$ denote the number of unknown parameters, and let $N_{\max}$ be the maximum number of the "1" elements in one row of all the matrices.

- When $N_P > N_M$, all the parameters in the same group are unidentifiable;
- When $N_P = N_M$, the parameters are globally identifiable if $N_{\max} = 1$, and locally identifiable if $N_{\max} > 1$;
- When $N_P < N_M$, the parameters are at least locally identifiable.

**Proof.** By definition of structural identifiability, the number of solutions to the identifiability equations is the key to verify parameter identifiability. According to the generation rule of identifiability matrices, $N_M$ identifiability matrices correspond to $N_M$ different identifiability equations. If the number of matrices is less than the number of unknown parameters (that is, $N_M < N_P$), the number of polynomial symbolic equations is then less than the number of unknown parameters. Because the identifiability equations are in the form of $\sigma_{ij} = \sum_{path_k} \prod_{edge_l} \theta_l$ and the order of $\theta_l$ is at most one in each monomial, this leads to an underdetermined system such that there exist an infinite number of solutions. Therefore, all the parameters are unidentifiable if the number of matrices is less than the number of parameters.

If the number of matrices is equal to the number of parameters in a group (i.e., $N_M = N_P$), then we have $N_P$ polynomial equations for real/complex variables with $N_P$ unknown parameters. According to the work of Garcia and Li [1], the number of solutions is equal to $q = \prod_{i=1}^{N_P} q_i$, where $q_i$ is the degree (the power of the highest order term) of equation $i$. If the maximum number $N_{\max}$ of the "1" elements in one row of all the matrices is equal to 1, then $q = 1$; this means that each parameter has a unique solution. In other words, every parameter is globally identifiable in this group. If the number $N_{\max}$ is greater than 1, then $q > 1$; this means there exist a finite number of solutions. In other words, every parameter is locally identifiable in the same group.

If the number of matrices is greater than the number of parameters in a group (i.e., $N_P < N_M$), then the equations corresponding to the matrices form an overdetermined polynomial system. For such a system, the number of solutions cannot be determined because even if we only change the constants in the system (e.g., constant coefficients), the number of solutions can be zero, one or a finite number, which needs to be analyzed case by case. However, from the parameter estimation point of view (i.e., the original problem becomes an optimization problem), there exist multiple local solutions [2-5]. Therefore, when $N_P < N_M$, the parameters are at least locally identifiable. ■

**Supporting Materials III —— Validation Using Benchmark Models**

**Benchmark Model 1**. See Fig. 6(b) in the manuscript.

**Benchmark Model 2**. Consider the mixed graph in Fig. S-1, which has been studied by Sullivant [1].

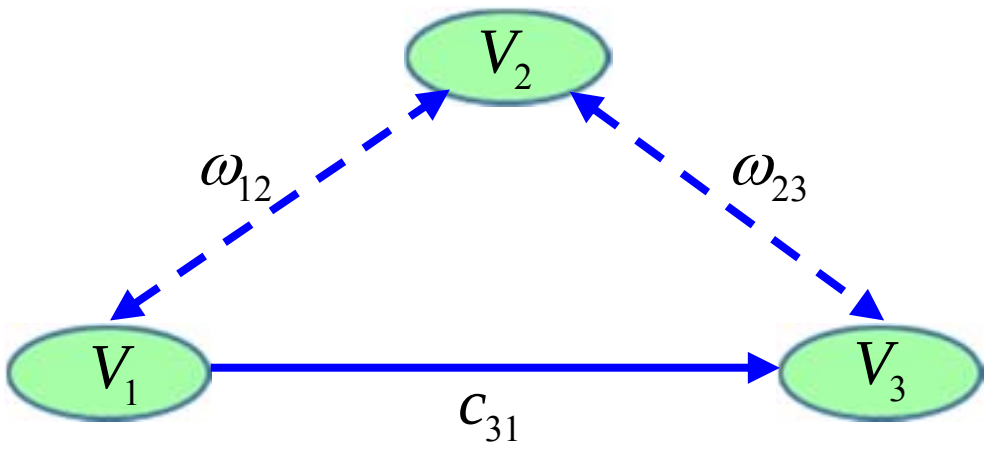

Figure S-1. A mixed graph with three nodes.

There are three parameters in this model, and the identifiability equations are as follows

$$\begin{cases} \sigma_{12} = \omega_{12} \\ \sigma_{13} = c_{31} \\ \sigma_{23} = \omega_{23} \end{cases} .$$

Let the column names of the identifiability matrices be $\{c_{31}, \omega_{12}, \omega_{23}\}$, and the identifiability matrices are generated from the identifiability equations as follows

$$\sigma_{12}\begin{bmatrix}0 & 1 & 0\end{bmatrix},$$

$$\sigma_{13}\begin{bmatrix}1 & 0 & 0\end{bmatrix},$$

$$\sigma_{23}\begin{bmatrix}0 & 0 & 1\end{bmatrix}.$$

Each matrix has only one row with only one "1" element and we cannot reduce the identifiability matrices. Therefore, all parameters are globally identifiable in this model.

**Benchmark Model 3**. Consider the mixed graph in Fig. S-2, which has been studied by Drton [2].

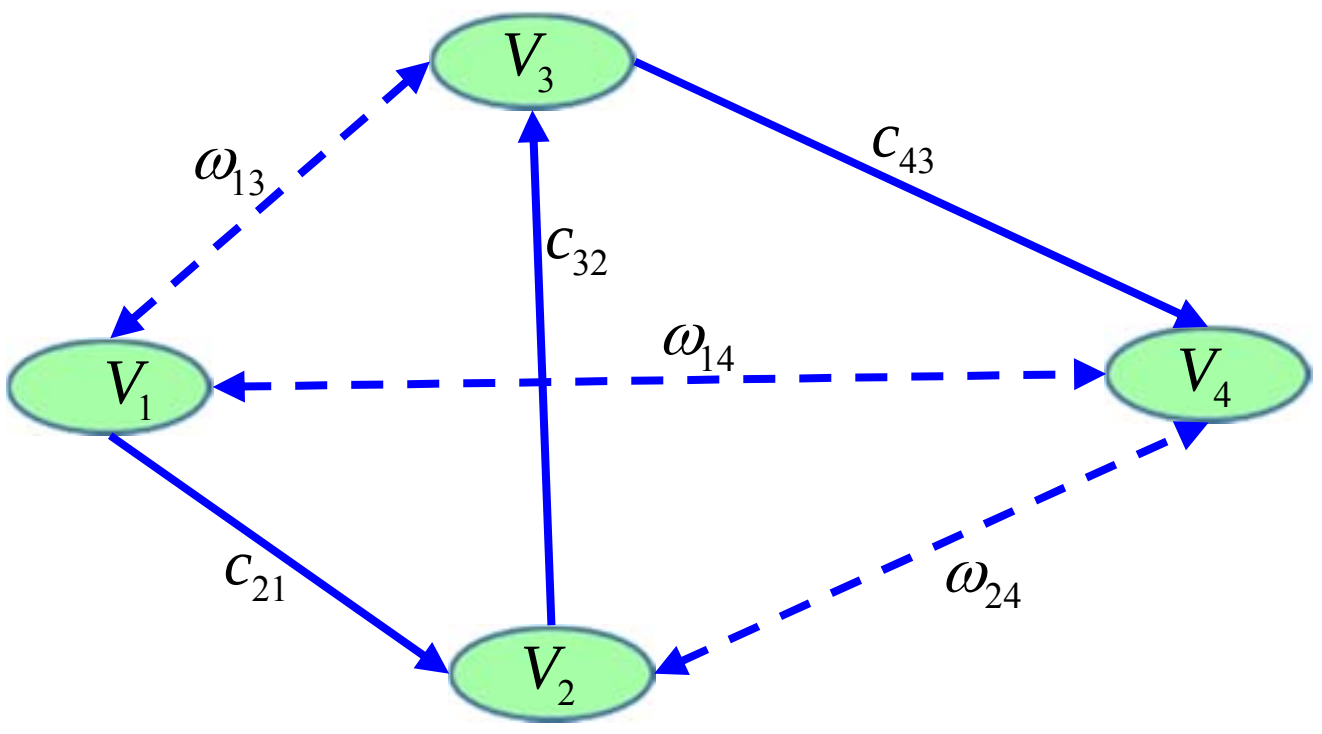

Figure S-2. A mixed graph with four nodes.

There are six parameters in this model and its identifiability equations are

$$\begin{cases} \sigma_{12} = c_{21} \\ \sigma_{13} = \omega_{13} + c_{21}c_{32} \\ \sigma_{14} = \omega_{14} + \omega_{13}c_{43} + c_{21}c_{32}c_{43} \\ \sigma_{23} = c_{32} + c_{21}\omega_{13} \\ \sigma_{24} = c_{32}c_{43} + \omega_{24} + c_{21}\omega_{14} + c_{21}\omega_{13}c_{43} \\ \sigma_{34} = c_{43} + c_{32}\omega_{24} + c_{32}c_{21}\omega_{14} \end{cases}.$$

The columns of the identifiability matrices correspond to $\{c_{21}, c_{32}, c_{43}, \omega_{13}, \omega_{14}, \omega_{24}\}$, respectively, and the identifiability matrices become

$$\sigma_{12}\begin{bmatrix} 1 & 0 & 0 & 0 & 0 & 0 \end{bmatrix},$$

$$\sigma_{13}\begin{bmatrix} 0 & 0 & 0 & 1 & 0 & 0 \\ 1 & 1 & 0 & 0 & 0 & 0 \end{bmatrix},$$

$$\sigma_{14}\begin{bmatrix} 0 & 0 & 0 & 0 & 1 & 0 \\ 0 & 0 & 1 & 1 & 0 & 0 \\ 1 & 1 & 1 & 0 & 0 & 0 \end{bmatrix},$$

$$\sigma_{23}\begin{bmatrix} 0 & 1 & 0 & 0 & 0 & 0 \\ 1 & 0 & 0 & 1 & 0 & 0 \end{bmatrix},$$

$$\sigma_{24}\begin{bmatrix} 0 & 1 & 1 & 0 & 0 & 0 \\ 0 & 0 & 0 & 0 & 0 & 1 \\ 1 & 0 & 0 & 0 & 1 & 0 \\ 1 & 0 & 1 & 1 & 0 & 0 \end{bmatrix},$$

$$\sigma_{34}\begin{bmatrix} 0 & 0 & 1 & 0 & 0 & 0 \\ 0 & 1 & 0 & 0 & 0 & 1 \\ 1 & 1 & 0 & 0 & 1 & 0 \end{bmatrix}.$$

Reduce matrix $\sigma_{14}$ with matrix $\sigma_{13}$, and also reduce matrix $\sigma_{24}$ with matrix $\sigma_{23}$, and we obtain

$$\sigma_{14}\begin{bmatrix} 0 & 0 & 0 & 0 & 1 & 0 \\ 0 & 0 & 1 & 0 & 0 & 0 \end{bmatrix},$$

$$\sigma_{24}\begin{bmatrix} 0 & 0 & 1 & 0 & 0 & 0 \\ 0 & 0 & 0 & 0 & 0 & 1 \\ 1 & 0 & 0 & 0 & 1 & 0 \end{bmatrix}.$$

Then reduce matrices $\sigma_{13}$, $\sigma_{23}$, $\sigma_{24}$ and $\sigma_{34}$ with matrix $\sigma_{12}$, matrix $\sigma_{23}$ with matrix $\sigma_{13}$, and matrix $\sigma_{24}$ with matrix $\sigma_{14}$, respectively, to obtain

$$\sigma_{13}\begin{bmatrix} 0 & 0 & 0 & 1 & 0 & 0 \\ 0 & 1 & 0 & 0 & 0 & 0 \end{bmatrix},$$

$$\sigma_{23}\begin{bmatrix} 0 & 1 & 0 & 0 & 0 & 0 \end{bmatrix},$$

$$\sigma_{24}\begin{bmatrix} 0 & 0 & 0 & 0 & 0 & 1 \end{bmatrix},$$

$$\sigma_{34}\begin{bmatrix} 0 & 0 & 1 & 0 & 0 & 0 \\ 0 & 1 & 0 & 0 & 0 & 1 \\ 0 & 1 & 0 & 0 & 1 & 0 \end{bmatrix}.$$

Now reduce $\sigma_{13}$ and $\sigma_{34}$ by $\sigma_{23}$ and $\sigma_{24}$ to obtain

$$\sigma_{13}\begin{bmatrix} 0 & 0 & 0 & 1 & 0 & 0 \end{bmatrix},$$

$$\sigma_{34}\begin{bmatrix} 0 & 0 & 1 & 0 & 0 & 0 \\ 0 & 0 & 0 & 0 & 1 & 0 \end{bmatrix}.$$

One can tell that the two matrices $\sigma_{14}$ and $\sigma_{34}$ are the same after row swapping. Finally, we reduce $\sigma_{14}$ with $\sigma_{34}$, and then reduce $\sigma_{34}$ with $\sigma_{14}$ to obtain $\sigma_{14}\begin{bmatrix} 0 & 0 & 0 & 0 & 1 & 0 \end{bmatrix}$ and $\sigma_{34}\begin{bmatrix} 0 & 0 & 1 & 0 & 0 & 0 \end{bmatrix}$. Now the reduction process is completed, and each matrix has only one row with only one "1" element in this row. Therefore, this model is globally identifiable.

**Benchmark Model 4**. Consider the mixed graph in Fig. S-3, which has also been studied by Drton [2].

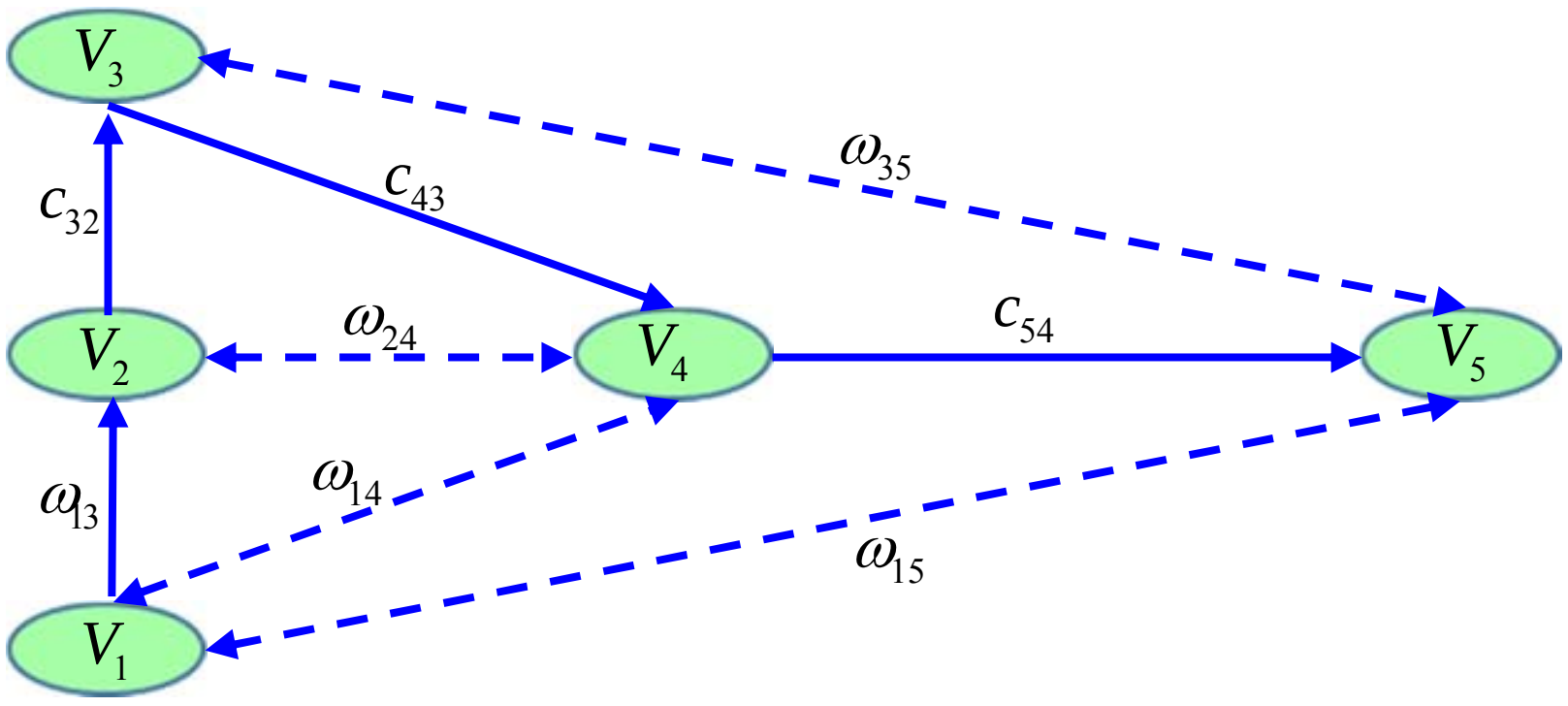

Figure S-3. A mixed graph with five nodes.

There are eight parameters in this model and the following identifiability equations can be generated

$$\begin{cases} \sigma_{12} = c_{21} \\ \sigma_{13} = c_{21}c_{32} \\ \sigma_{14} = \omega_{14} + c_{21}c_{32}c_{43} \\ \sigma_{15} = \omega_{15} + \omega_{14}c_{54} + c_{21}c_{32}c_{43}c_{54} \\ \sigma_{23} = c_{32} \\ \sigma_{24} = c_{32}c_{43} + \omega_{24} + \omega_{14}c_{21} \\ \sigma_{25} = \omega_{24}c_{54} + c_{32}c_{43}c_{54} + c_{21}\omega_{14}c_{54} + c_{21}\omega_{15} \\ \sigma_{34} = c_{43} + c_{32}\omega_{24} + c_{32}c_{21}\omega_{14} \\ \sigma_{35} = \omega_{35} + c_{43}c_{54} + c_{32}\omega_{24}c_{54} + c_{32}c_{21}\omega_{14}c_{54} + c_{32}c_{21}\omega_{15} \\ \sigma_{45} = c_{54} + c_{43}\omega_{35} + c_{43}c_{32}c_{21}\omega_{15} \end{cases}.$$

The columns of the identifiability matrices correspond to $\{c_{21}, c_{32}, c_{43}, c_{54}, \omega_{14}, \omega_{15}, \omega_{24}, \omega_{35}\}$, respectively, and the identifiability matrices can be derived as follows

$$\sigma_{12}\begin{bmatrix}1 & 0 & 0 & 0 & 0 & 0 & 0 & 0\end{bmatrix},$$

$$\sigma_{13}\begin{bmatrix}1 & 1 & 0 & 0 & 0 & 0 & 0 & 0\end{bmatrix},$$

$$\sigma_{14}\begin{bmatrix}0 & 0 & 0 & 0 & 1 & 0 & 0 & 0 \\ 1 & 1 & 1 & 0 & 0 & 0 & 0 & 0\end{bmatrix},$$

$$\sigma_{15}\begin{bmatrix}0 & 0 & 0 & 0 & 0 & 1 & 0 & 0 \\ 0 & 0 & 0 & 1 & 1 & 0 & 0 & 0 \\ 1 & 1 & 1 & 1 & 0 & 0 & 0 & 0\end{bmatrix},$$

$$\sigma_{23}\begin{bmatrix}0 & 1 & 0 & 0 & 0 & 0 & 0 & 0\end{bmatrix},$$

$$\sigma_{24}\begin{bmatrix}0&1&1&0&0&0&0&0\\0&0&0&0&0&0&1&0\\1&0&0&0&1&0&0&0\end{bmatrix},$$

$$\sigma_{25}\begin{bmatrix}0&0&0&1&0&0&1&0\\0&1&1&1&0&0&0&0\\1&0&0&1&1&0&0&0\\1&0&0&0&0&1&0&0\end{bmatrix},$$

$$\sigma_{34}\begin{bmatrix}0&0&1&0&0&0&0&0\\0&1&0&0&0&0&1&0\\1&1&0&0&1&0&0&0\end{bmatrix},$$

$$\sigma_{35}\begin{bmatrix}0&0&0&0&0&0&0&1\\0&0&1&1&0&0&0&0\\0&1&0&1&0&0&1&0\\1&1&0&1&1&0&0&0\\1&1&0&0&0&1&0&0\end{bmatrix},$$

$$\sigma_{45}\begin{bmatrix}0&0&0&1&0&0&0&0\\0&0&1&0&0&0&0&1\\1&1&1&0&0&1&0&0\end{bmatrix}.$$

Reduce matrices $\sigma_{13}$, $\sigma_{14}$, $\sigma_{15}$, $\sigma_{24}$, $\sigma_{25}$, $\sigma_{34}$, $\sigma_{35}$ and $\sigma_{45}$ with matrices $\sigma_{12}$ and $\sigma_{23}$ to obtain

$$\sigma_{13}\begin{bmatrix}0&1&0&0&0&0&0&0\end{bmatrix},$$

$$\sigma_{14}\begin{bmatrix}0&0&0&0&1&0&0&0\\0&0&1&0&0&0&0&0\end{bmatrix},$$

$$\sigma_{15}\begin{bmatrix}0&0&0&0&0&1&0&0\\0&0&0&1&1&0&0&0\\0&0&1&1&0&0&0&0\end{bmatrix},$$

$$\sigma_{24}\begin{bmatrix}0&0&1&0&0&0&0&0\\0&0&0&0&0&0&1&0\\0&0&0&0&1&0&0&0\end{bmatrix},$$

$$\sigma_{25}\begin{bmatrix}0&0&0&1&0&0&1&0\\0&0&1&1&0&0&0&0\\0&0&0&1&1&0&0&0\\0&0&0&0&0&1&0&0\end{bmatrix},$$

$$\sigma_{34}\begin{bmatrix}0&0&1&0&0&0&0&0\\0&0&0&0&0&0&1&0\\0&0&0&0&1&0&0&0\end{bmatrix},$$

$$\sigma_{35}\begin{bmatrix}0&0&0&0&0&0&0&1\\0&0&1&1&0&0&0&0\\0&0&0&1&0&0&1&0\\0&0&0&1&1&0&0&0\\0&0&0&0&0&1&0&0\end{bmatrix},$$

$$\sigma_{45}\begin{bmatrix}0&0&0&1&0&0&0&0\\0&0&1&0&0&0&0&1\\0&0&1&0&0&1&0&0\end{bmatrix}.$$

Then reduce matrix $\sigma_{15}$ with matrix $\sigma_{14}$, and reduce matrices $\sigma_{25}$, $\sigma_{34}$, and $\sigma_{35}$ with matrix $\sigma_{24}$, to obtain

$$\sigma_{15}\begin{bmatrix}0&0&0&0&0&1&0&0\\0&0&0&1&0&0&0&0\end{bmatrix},$$

$$\sigma_{25}\begin{bmatrix}0&0&0&1&0&0&0&0\\0&0&0&0&0&1&0&0\end{bmatrix},$$

$$\sigma_{34}\begin{bmatrix}0&0&1&0&0&0&0&0\end{bmatrix},$$

$$\sigma_{35}\begin{bmatrix}0&0&0&0&0&0&0&1\\0&0&0&1&0&0&0&0\\0&0&0&0&0&1&0&0\end{bmatrix}.$$

Now reduce matrices $\sigma_{14}$, $\sigma_{24}$ and $\sigma_{45}$ with matrix $\sigma_{34}$, and reduce matrices $\sigma_{25}$ and $\sigma_{35}$ with matrix $\sigma_{15}$, and we get

$$\sigma_{14}\begin{bmatrix}0&0&0&0&1&0&0&0\end{bmatrix},$$

$$\sigma_{24}\begin{bmatrix}0&0&0&0&0&0&1&0\\0&0&0&0&1&0&0&0\end{bmatrix},$$

$\sigma_{25}\begin{bmatrix}0 & 0 & 0 & 1 & 0 & 0 & 0 & 0\end{bmatrix}$,

$\sigma_{35}\begin{bmatrix}0 & 0 & 0 & 0 & 0 & 0 & 0 & 1\end{bmatrix}$,

$$\sigma_{45}\begin{bmatrix}0 & 0 & 0 & 1 & 0 & 0 & 0 & 0\\ 0 & 0 & 0 & 0 & 0 & 0 & 0 & 1\\ 0 & 0 & 0 & 0 & 0 & 1 & 0 & 0\end{bmatrix}.$$

Further reduce matrix $\sigma_{45}$ with matrix $\sigma_{15}$ to obtain $\sigma_{45}\begin{bmatrix}0 & 0 & 0 & 0 & 0 & 0 & 0 & 1\end{bmatrix}$, and matrix $\sigma_{15}$ with matrix $\sigma_{25}$ to obtain $\sigma_{15}\begin{bmatrix}0 & 0 & 0 & 0 & 0 & 1 & 0 & 0\end{bmatrix}$, and matrix $\sigma_{24}$ with matrix $\sigma_{14}$ to obtain $\sigma_{24}\begin{bmatrix}0 & 0 & 0 & 0 & 0 & 0 & 1 & 0\end{bmatrix}$. Now each matrix has only one row that has only one "1" element. Therefore, this model is globally identifiable. Note that there exist two groups of matrices that are the same: 1) $\sigma_{13}$ and $\sigma_{23}$; 2) $\sigma_{35}$ and $\sigma_{45}$, and it indicates that there exist two redundant identifiability equations among the original identifiability equations.

**Benchmark Model 5**. Consider the mixed graph in Fig. S-4, which has been studied by Kline [3]. Note that the measurement model of the original graph is ignored here because it is not the focus of this study.

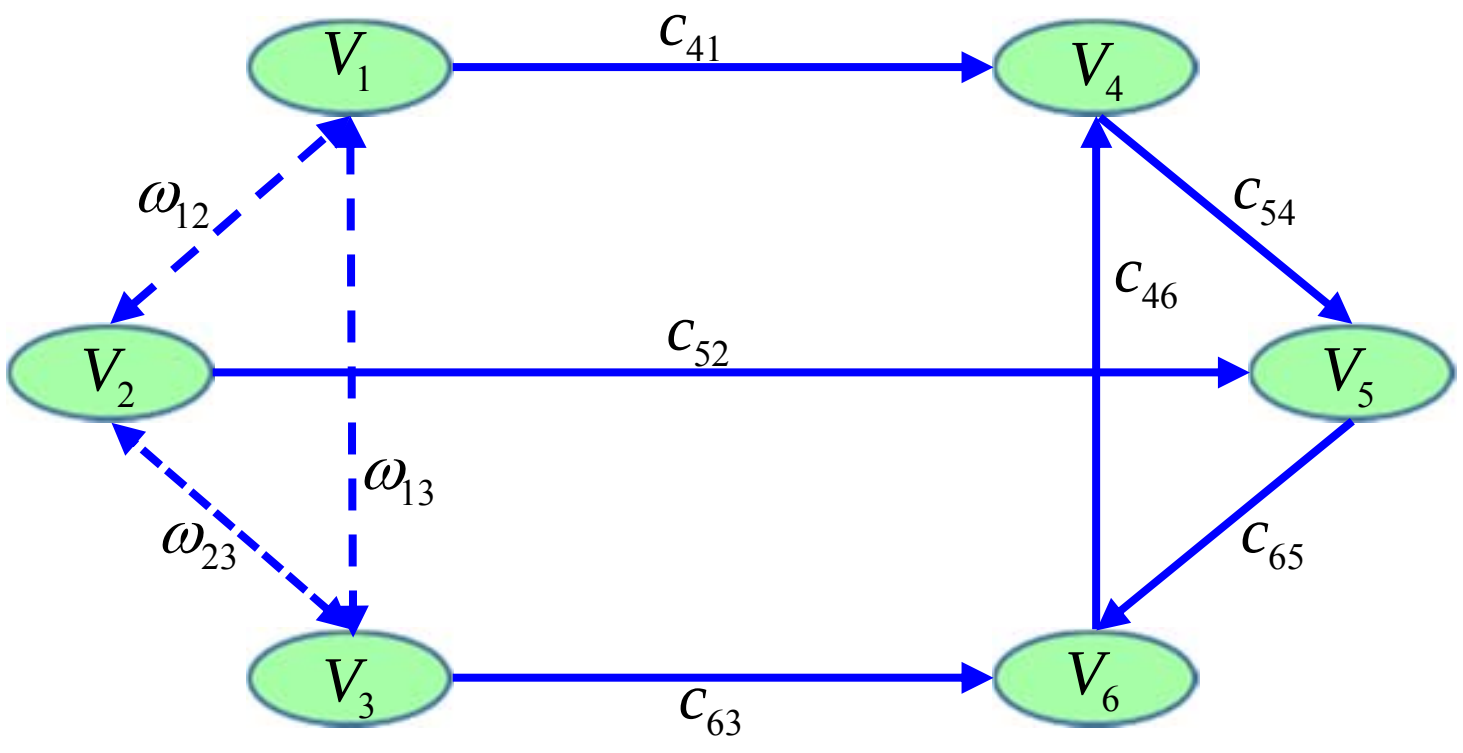

Figure S-4. A mixed graph with six nodes and circles.

There are nine parameters in this model and its identifiability equations are as follows

$$
\begin{cases}
\sigma_{12} = \omega_{12} \\
\sigma_{13} = \omega_{13} \\
\sigma_{14} = c_{41} + \omega_{12}c_{52}c_{65}c_{46} + \omega_{13}c_{63}c_{46} \\
\sigma_{15} = c_{41}c_{54} + \omega_{12}c_{52} + \omega_{13}c_{63}c_{46}c_{54} \\
\sigma_{16} = c_{41}c_{54}c_{65} + \omega_{12}c_{52}c_{65} + \omega_{13}c_{63} \\
\sigma_{23} = \omega_{23} \\
\sigma_{24} = \omega_{12}c_{41} + c_{52}c_{65}c_{46} + \omega_{23}c_{63}c_{46} \\
\sigma_{25} = c_{52} + \omega_{12}c_{41}c_{54} + \omega_{23}c_{63}c_{46}c_{54} \\
\sigma_{26} = c_{52}c_{65} + \omega_{23}c_{63} + \omega_{12}c_{41}c_{54}c_{65} \\
\sigma_{34} = c_{63}c_{46} + \omega_{23}c_{52}c_{65}c_{46} + \omega_{13}c_{41} \\
\sigma_{35} = c_{63}c_{46}c_{54} + \omega_{13}c_{41}c_{54} + \omega_{23}c_{52} \\
\sigma_{36} = c_{63} + \omega_{23}c_{52}c_{65} + \omega_{13}c_{41}c_{54}c_{65} \\
\sigma_{45} = c_{54} + c_{41}\omega_{12}c_{52} \\
\sigma'_{45} = c_{65}c_{46} + c_{41}\omega_{12}c_{52} \\
\sigma_{46} = c_{54}c_{65} + c_{41}\omega_{13}c_{63} + c_{41}\omega_{12}c_{52}c_{65} \\
\sigma'_{46} = c_{46} + c_{41}\omega_{13}c_{63} \\
\sigma_{56} = c_{65} + c_{52}\omega_{23}c_{63} \\
\sigma'_{56} = c_{46}c_{54} + c_{52}\omega_{23}c_{63} + c_{63}\omega_{13}c_{41}c_{54}
\end{cases}
.
$$

The columns of the identifiability matrices correspond to $\{c_{41}, c_{52}, c_{63}, c_{46}, c_{54}, c_{65}, \omega_{12}, \omega_{13}, \omega_{23}\}$, respectively, and the identifiability matrices can be generated as follows

$$\sigma_{12}\begin{bmatrix} 0 & 0 & 0 & 0 & 0 & 0 & 1 & 0 & 0 \end{bmatrix},$$

$$\sigma_{13}\begin{bmatrix} 0 & 0 & 0 & 0 & 0 & 0 & 0 & 1 & 0 \end{bmatrix},$$

$$\sigma_{14}\begin{bmatrix} 1 & 0 & 0 & 0 & 0 & 0 & 0 & 0 & 0 \\ 0 & 1 & 0 & 1 & 0 & 1 & 1 & 0 & 0 \\ 0 & 0 & 1 & 1 & 0 & 0 & 0 & 1 & 0 \end{bmatrix},$$

$$\sigma_{15}\begin{bmatrix} 1 & 0 & 0 & 0 & 1 & 0 & 0 & 0 & 0 \\ 0 & 1 & 0 & 0 & 0 & 0 & 1 & 0 & 0 \\ 0 & 0 & 1 & 1 & 1 & 0 & 0 & 1 & 0 \end{bmatrix},$$

$$\sigma_{16}\begin{bmatrix} 1 & 0 & 0 & 0 & 1 & 1 & 0 & 0 & 0 \\ 0 & 1 & 0 & 0 & 0 & 1 & 1 & 0 & 0 \\ 0 & 0 & 1 & 0 & 0 & 0 & 0 & 1 & 0 \end{bmatrix},$$

$$\sigma_{23}\begin{bmatrix} 0 & 0 & 0 & 0 & 0 & 0 & 0 & 0 & 1 \end{bmatrix},$$

$$\sigma_{24}\begin{bmatrix}1&0&0&0&0&0&1&0&0\\0&1&0&1&0&1&0&0&0\\0&0&1&1&0&0&0&0&1\end{bmatrix},$$

$$\sigma_{25}\begin{bmatrix}0&1&0&0&0&0&0&0&0\\1&0&0&0&1&0&1&0&0\\0&0&1&1&1&0&0&0&1\end{bmatrix},$$

$$\sigma_{26}\begin{bmatrix}0&1&0&0&0&1&0&0&0\\0&0&1&0&0&0&0&0&1\\1&0&0&0&1&1&1&0&0\end{bmatrix},$$

$$\sigma_{34}\begin{bmatrix}0&0&1&1&0&0&0&0&0\\0&1&0&1&0&1&0&0&1\\1&0&0&0&0&0&0&1&0\end{bmatrix},$$

$$\sigma_{35}\begin{bmatrix}0&0&1&1&1&0&0&0&0\\1&0&0&0&1&0&0&1&0\\0&1&0&0&0&0&0&0&1\end{bmatrix},$$

$$\sigma_{36}\begin{bmatrix}0&0&1&0&0&0&0&0&0\\0&1&0&0&0&1&0&0&1\\1&0&0&0&1&1&0&1&0\end{bmatrix},$$

$$\sigma_{45}\begin{bmatrix}0&0&0&0&1&0&0&0&0\\1&1&0&0&0&0&1&0&0\end{bmatrix},$$

$$\sigma_{45}^{'}\begin{bmatrix}0&0&0&1&0&1&0&0&0\\1&1&0&0&0&0&1&0&0\end{bmatrix},$$

$$\sigma_{46}\begin{bmatrix}0&0&0&1&1&0&0&0&0\\1&0&1&0&0&0&0&1&0\\1&1&0&0&0&1&1&0&0\end{bmatrix},$$

$$\sigma_{46}^{'}\begin{bmatrix}0&0&0&1&0&0&0&0&0\\1&0&1&0&0&0&0&1&0\end{bmatrix},$$

$$\sigma_{56}\begin{bmatrix}0&0&0&0&0&1&0&0&0\\0&1&1&0&0&0&0&0&1\end{bmatrix},$$

$$\sigma_{56}^{'}\begin{bmatrix}0&0&0&1&1&0&0&0&0\\0&1&1&0&0&0&0&0&1\\1&0&1&0&1&0&0&1&0\end{bmatrix}.$$

Reduce matrices $\sigma_{14}$, $\sigma_{15}$, $\sigma_{16}$, $\sigma_{24}$, $\sigma_{25}$, $\sigma_{26}$, $\sigma_{34}$, $\sigma_{35}$, $\sigma_{36}$, $\sigma_{45}$, $\sigma_{54}$, $\sigma_{46}$, $\sigma_{64}$, $\sigma_{56}$ and $\sigma_{65}$ with matrices $\sigma_{12}$, $\sigma_{13}$ and $\sigma_{23}$, and obtain

$$\sigma_{14}\begin{bmatrix} 1 & 0 & 0 & 0 & 0 & 0 & 0 & 0 & 0 \\ 0 & 1 & 0 & 1 & 0 & 1 & 0 & 0 & 0 \\ 0 & 0 & 1 & 1 & 0 & 0 & 0 & 0 & 0 \end{bmatrix},$$

$$\sigma_{15}\begin{bmatrix} 1 & 0 & 0 & 0 & 1 & 0 & 0 & 0 & 0 \\ 0 & 1 & 0 & 0 & 0 & 0 & 0 & 0 & 0 \\ 0 & 0 & 1 & 1 & 1 & 0 & 0 & 0 & 0 \end{bmatrix},$$

$$\sigma_{16}\begin{bmatrix} 1 & 0 & 0 & 0 & 1 & 1 & 0 & 0 & 0 \\ 0 & 1 & 0 & 0 & 0 & 1 & 0 & 0 & 0 \\ 0 & 0 & 1 & 0 & 0 & 0 & 0 & 0 & 0 \end{bmatrix},$$

$$\sigma_{24}\begin{bmatrix} 1 & 0 & 0 & 0 & 0 & 0 & 0 & 0 & 0 \\ 0 & 1 & 0 & 1 & 0 & 1 & 0 & 0 & 0 \\ 0 & 0 & 1 & 1 & 0 & 0 & 0 & 0 & 0 \end{bmatrix},$$

$$\sigma_{25}\begin{bmatrix} 0 & 1 & 0 & 0 & 0 & 0 & 0 & 0 & 0 \\ 1 & 0 & 0 & 0 & 1 & 0 & 0 & 0 & 0 \\ 0 & 0 & 1 & 1 & 1 & 0 & 0 & 0 & 0 \end{bmatrix},$$

$$\sigma_{26}\begin{bmatrix} 0 & 1 & 0 & 0 & 0 & 1 & 0 & 0 & 0 \\ 0 & 0 & 1 & 0 & 0 & 0 & 0 & 0 & 0 \\ 1 & 0 & 0 & 0 & 1 & 1 & 0 & 0 & 0 \end{bmatrix},$$

$$\sigma_{34}\begin{bmatrix} 0 & 0 & 1 & 1 & 0 & 0 & 0 & 0 & 0 \\ 0 & 1 & 0 & 1 & 0 & 1 & 0 & 0 & 0 \\ 1 & 0 & 0 & 0 & 0 & 0 & 0 & 0 & 0 \end{bmatrix},$$

$$\sigma_{35}\begin{bmatrix} 0 & 0 & 1 & 1 & 1 & 0 & 0 & 0 & 0 \\ 1 & 0 & 0 & 0 & 1 & 0 & 0 & 0 & 0 \\ 0 & 1 & 0 & 0 & 0 & 0 & 0 & 0 & 0 \end{bmatrix},$$

$$\sigma_{36}\begin{bmatrix} 0 & 0 & 1 & 0 & 0 & 0 & 0 & 0 & 0 \\ 0 & 1 & 0 & 0 & 0 & 1 & 0 & 0 & 0 \\ 1 & 0 & 0 & 0 & 1 & 1 & 0 & 0 & 0 \end{bmatrix},$$

$$\sigma_{45}\begin{bmatrix} 0 & 0 & 0 & 0 & 1 & 0 & 0 & 0 & 0 \\ 1 & 1 & 0 & 0 & 0 & 0 & 0 & 0 & 0 \end{bmatrix},$$

$$\sigma_{45}^{'}\begin{bmatrix}0&0&0&1&0&1&0&0&0\\1&1&0&0&0&0&0&0&0\end{bmatrix},$$

$$\sigma_{46}\begin{bmatrix}0&0&0&0&1&1&0&0&0\\1&0&1&0&0&0&0&0&0\\1&1&0&0&0&1&0&0&0\end{bmatrix},$$

$$\sigma_{46}^{'}\begin{bmatrix}0&0&0&1&0&0&0&0&0\\1&0&1&0&0&0&0&0&0\end{bmatrix},$$

$$\sigma_{56}\begin{bmatrix}0&0&0&0&0&1&0&0&0\\0&1&1&0&0&0&0&0&0\end{bmatrix},$$

$$\sigma_{56}^{'}\begin{bmatrix}0&0&0&1&1&0&0&0&0\\0&1&1&0&0&0&0&0&0\\1&0&1&0&1&0&0&0&0\end{bmatrix}.$$

Now there exist three equivalent groups of matrices after row swapping: 1) $\sigma_{14}$, $\sigma_{24}$ and $\sigma_{34}$; 2) $\sigma_{15}$, $\sigma_{25}$ and $\sigma_{35}$; 3) $\sigma_{16}$, $\sigma_{26}$ and $\sigma_{36}$. Reduce the three groups of matrices, respectively, and we obtain

$$\sigma_{14}\begin{bmatrix}1&0&0&0&0&0&0&0&0\end{bmatrix},$$

$$\sigma_{15}\begin{bmatrix}1&0&0&0&1&0&0&0&0\end{bmatrix},$$

$$\sigma_{16}\begin{bmatrix}1&0&0&0&1&1&0&0&0\end{bmatrix},$$

$$\sigma_{24}\begin{bmatrix}0&1&0&1&0&1&0&0&0\end{bmatrix},$$

$$\sigma_{25}\begin{bmatrix}0&1&0&0&0&0&0&0&0\end{bmatrix},$$

$$\sigma_{26}\begin{bmatrix}0&1&0&0&0&1&0&0&0\end{bmatrix},$$

$$\sigma_{34}\begin{bmatrix}0&0&1&1&0&0&0&0&0\end{bmatrix},$$

$$\sigma_{35}\begin{bmatrix}0&0&1&1&1&0&0&0&0\end{bmatrix},$$

$$\sigma_{36}\begin{bmatrix}0&0&1&0&0&0&0&0&0\end{bmatrix}.$$

Reduce matrices $\sigma_{15}$, $\sigma_{16}$, $\sigma_{24}$, $\sigma_{26}$, $\sigma_{34}$, $\sigma_{35}$, $\sigma_{45}$, $\sigma_{45}^{'}$, $\sigma_{46}$, $\sigma_{46}^{'}$, $\sigma_{56}$ and $\sigma_{56}^{'}$ with matrices $\sigma_{14}$, $\sigma_{25}$ and $\sigma_{36}$, and get the following results

$\sigma_{15}\begin{bmatrix}0 & 0 & 0 & 0 & 1 & 0 & 0 & 0 & 0\end{bmatrix}$,

$\sigma_{16}\begin{bmatrix}0 & 0 & 0 & 0 & 1 & 1 & 0 & 0 & 0\end{bmatrix}$,

$\sigma_{24}\begin{bmatrix}0 & 0 & 0 & 1 & 0 & 1 & 0 & 0 & 0\end{bmatrix}$,

$\sigma_{26}\begin{bmatrix}0 & 0 & 0 & 0 & 0 & 1 & 0 & 0 & 0\end{bmatrix}$,

$\sigma_{34}\begin{bmatrix}0 & 0 & 0 & 1 & 0 & 0 & 0 & 0 & 0\end{bmatrix}$,

$\sigma_{35}\begin{bmatrix}0 & 0 & 0 & 1 & 1 & 0 & 0 & 0 & 0\end{bmatrix}$,

$\sigma_{45}\begin{bmatrix}0 & 0 & 0 & 0 & 1 & 0 & 0 & 0 & 0\end{bmatrix}$,

$\sigma_{45}^{'}\begin{bmatrix}0 & 0 & 0 & 1 & 0 & 1 & 0 & 0 & 0\end{bmatrix}$,

$$\sigma_{46}\begin{bmatrix}0 & 0 & 0 & 0 & 1 & 1 & 0 & 0 & 0\\ 0 & 0 & 0 & 0 & 0 & 1 & 0 & 0 & 0\end{bmatrix},$$

$\sigma_{46}^{'}\begin{bmatrix}0 & 0 & 0 & 1 & 0 & 0 & 0 & 0 & 0\end{bmatrix}$,

$\sigma_{56}\begin{bmatrix}0 & 0 & 0 & 0 & 0 & 1 & 0 & 0 & 0\end{bmatrix}$,

$$\sigma_{56}^{'}\begin{bmatrix}0 & 0 & 0 & 1 & 1 & 0 & 0 & 0 & 0\\ 0 & 0 & 0 & 0 & 1 & 0 & 0 & 0 & 0\end{bmatrix}.$$

Further reduce matrices $\sigma_{16}$, $\sigma_{35}$, $\sigma_{46}$ and $\sigma_{56}^{'}$ with matrix $\sigma_{15}$, and also reduce matrices $\sigma_{24}$ and $\sigma_{45}^{'}$ with matrix $\sigma_{26}$ to obtain

$\sigma_{16}\begin{bmatrix}0 & 0 & 0 & 0 & 0 & 1 & 0 & 0 & 0\end{bmatrix}$,

$\sigma_{24}\begin{bmatrix}0 & 0 & 0 & 1 & 0 & 0 & 0 & 0 & 0\end{bmatrix}$,

$\sigma_{35}\begin{bmatrix}0 & 0 & 0 & 1 & 0 & 0 & 0 & 0 & 0\end{bmatrix}$,

$\sigma_{45}^{'}\begin{bmatrix}0 & 0 & 0 & 1 & 0 & 0 & 0 & 0 & 0\end{bmatrix}$,

$\sigma_{46}\begin{bmatrix}0 & 0 & 0 & 0 & 0 & 1 & 0 & 0 & 0\end{bmatrix}$,

$\sigma_{56}^{'}\begin{bmatrix}0 & 0 & 0 & 1 & 0 & 0 & 0 & 0 & 0\end{bmatrix}$.

Now the reduction process is finished. Each of the remaining matrices has only one row with only one “1” element. Therefore, this model is globally identifiable. Note that there exist three equivalent groups of matrices: 1) $\sigma_{15}$ and $\sigma_{45}$; 2) $\sigma_{16}$, $\sigma_{26}$, $\sigma_{46}$ and $\sigma_{56}$; 3) $\sigma_{24}$, $\sigma_{34}$, $\sigma_{35}$, $\sigma_{45}^{'}$, $\sigma_{46}^{'}$ and $\sigma_{56}^{'}$, and it suggests that there exist nine redundant identifiability equations among the original identifiability equations.

**Benchmark Model 6**. Consider the mixed graph in Fig. S-5, which has been studied by Brito and Pearl [4]. Note that there are two different edges (one directed and one undirected) from node $V_3$ to node $V_5$ and from node $V_3$ to node $V_5$, respectively.

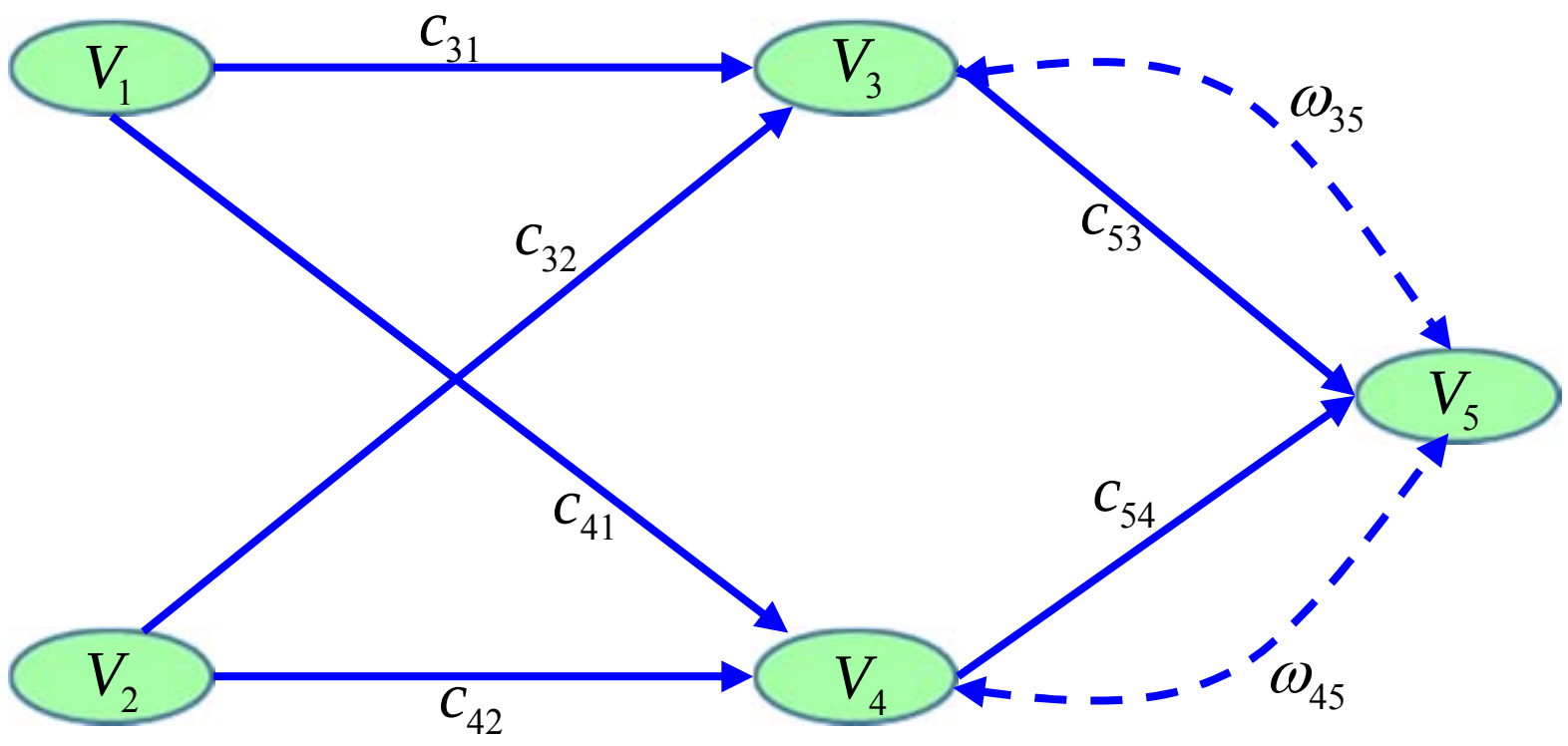

Figure S-5. A mixed graph with five nodes and repeated edges.

There are eight parameters in this model and its identifiability equations are given as follows

$$\begin{cases} \sigma_{13} = c_{31} \\ \sigma_{23} = c_{32} \\ \sigma_{14} = c_{41} \\ \sigma_{24} = c_{42} \\ \sigma_{15} = c_{31}c_{53} + c_{41}c_{54} \\ \sigma_{25} = c_{42}c_{54} + c_{32}c_{53} \\ \sigma_{34} = c_{32}c_{42} + c_{31}c_{41} \\ \sigma_{35} = c_{53} + \omega_{35} + c_{31}c_{41}c_{54} + c_{32}c_{42}c_{54} \\ \sigma_{45} = c_{54} + \omega_{45} + c_{31}c_{41}c_{53} + c_{32}c_{42}c_{53} \end{cases}.$$

The columns of the identifiability matrices correspond to $\{c_{31}, c_{32}, c_{41}, c_{42}, c_{53}, c_{54}, \omega_{35}, \omega_{45}\}$, respectively, and then the identifiability matrices can be derived as follows

$\sigma_{13}\begin{bmatrix}1 & 0 & 0 & 0 & 0 & 0 & 0 & 0\end{bmatrix}$,

$\sigma_{23}\begin{bmatrix}0 & 1 & 0 & 0 & 0 & 0 & 0 & 0\end{bmatrix}$,

$$\sigma_{14}\begin{bmatrix}0&0&1&0&0&0&0&0\end{bmatrix},$$

$$\sigma_{24}\begin{bmatrix}0&0&0&1&0&0&0&0\end{bmatrix},$$

$$\sigma_{15}\begin{bmatrix}1&0&0&0&1&0&0&0\\0&0&1&0&0&1&0&0\end{bmatrix},$$

$$\sigma_{25}\begin{bmatrix}0&0&0&1&0&1&0&0\\0&1&0&0&1&0&0&0\end{bmatrix},$$

$$\sigma_{34}\begin{bmatrix}0&1&0&1&0&0&0&0\\1&0&1&0&0&0&0&0\end{bmatrix},$$

$$\sigma_{35}\begin{bmatrix}0&0&0&0&1&0&0&0\\0&0&0&0&0&0&1&0\\1&0&1&0&0&1&0&0\\0&1&0&1&0&1&0&0\end{bmatrix},$$

$$\sigma_{45}\begin{bmatrix}0&0&0&0&0&1&0&0\\0&0&0&0&0&0&0&1\\1&0&1&0&1&0&0&0\\0&1&0&1&1&0&0&0\end{bmatrix}.$$

Reduce matrices $\sigma_{15}$, $\sigma_{25}$, $\sigma_{34}$, $\sigma_{35}$ and $\sigma_{45}$ with matrices $\sigma_{13}$, $\sigma_{23}$, $\sigma_{14}$ and $\sigma_{24}$, and get the following matrices

$$\sigma_{15}\begin{bmatrix}0&0&0&0&1&0&0&0\\0&0&0&0&0&1&0&0\end{bmatrix},$$

$$\sigma_{25}\begin{bmatrix}0&0&0&0&0&1&0&0\\0&0&0&0&1&0&0&0\end{bmatrix},$$

$$\sigma_{34}\begin{bmatrix}0&0&0&1&0&0&0&0\end{bmatrix},$$

$$\sigma_{35}\begin{bmatrix}0&0&0&0&1&0&0&0\\0&0&0&0&0&0&1&0\\0&0&0&0&0&1&0&0\end{bmatrix},$$

$$\sigma_{45}\begin{bmatrix}0&0&0&0&0&1&0&0\\0&0&0&0&0&0&0&1\\0&0&0&0&1&0&0&0\end{bmatrix}.$$

Reduce matrices $\sigma_{25}$, $\sigma_{35}$ and $\sigma_{45}$ with matrix $\sigma_{15}$, and get

$\sigma_{25}\begin{bmatrix}0 & 0 & 0 & 0 & 0 & 1 & 0 & 0\end{bmatrix}$,

$\sigma_{35}\begin{bmatrix}0 & 0 & 0 & 0 & 0 & 0 & 1 & 0\end{bmatrix}$,

$\sigma_{45}\begin{bmatrix}0 & 0 & 0 & 0 & 0 & 0 & 0 & 1\end{bmatrix}$.

Reduce matrix $\sigma_{15}$ with matrix $\sigma_{25}$, and get $\sigma_{15}\begin{bmatrix}0 & 0 & 0 & 0 & 1 & 0 & 0 & 0\end{bmatrix}$. Now the reduction process is done, and each matrix has only one row with only one "1" element. Therefore, this model is globally identifiable. Note that there exists one redundant identifiability equations among all the original identifiability equations, because there exists a group of the same matrices: $\sigma_{24}$ and $\sigma_{34}$.